\documentclass[showpacs,twocolumn,amsmath,amssymb]{revtex4}

\usepackage{graphicx}
\usepackage{algorithm}
\usepackage{algorithmic}

 \usepackage{amsmath}
 \usepackage{amssymb}
 \DeclareMathOperator*{\argmax}{argmax}

\makeatother

\begin{document}

\title{Spin glass phase transitions in the random feedback vertex set problem}

\author{Shao-Meng Qin$^{1,3}$\footnote{Email: {\tt qsminside@163.com}.},
  Ying Zeng$^{2}$,
  and Hai-Jun Zhou$^{1,4}$\footnote{Corresponding author. Email: 
    {\tt zhouhj@itp.ac.cn}.}
}
\affiliation{
  $^1$Institute of Theoretical Physics, Chinese Academy of Sciences,
  Zhong-Guan-Cun East Road 55, Beijing 100190, China\\
  $^2$
  Beijing National Laboratory for Condensed Matter Physics and
  Key Laboratory of Soft Matter Physics, Institute of Physics,
  Chinese Academy of Sciences, Beijing 100190, China \\
  $^3$College of Science, Civil Aviation University of China, Tianjin 300300,
  China \\
  $^4$School of Physical Sciences, University of Chinese Academy of Sciences,
  Beijing 100049, China
}

\date{February 07, 2016}

\begin{abstract}
  A feedback vertex set (FVS) of an undirected graph contains vertices from
  every cycle of this graph. Constructing a FVS of sufficiently small
  cardinality is very difficult in the worst cases, but for random graphs this
  problem can be efficiently solved after converting it into an appropriate
  spin glass model [H.-J.~Zhou, Eur.~Phys.~J.~B {\bf 86} (2013) 455].
  In the present work we study the local stability and the phase transition
  properties of this spin glass model on random graphs.
  For both regular random graphs and Erd\"os-R\'enyi graphs we determine the
  inverse temperature $\beta_l$ at which the replica-symmetric mean field
  theory loses its local stability, the inverse temperature $\beta_d$ of
  the dynamical (clustering) phase transition, and the inverse temperature
  $\beta_c$ of the static (condensation) phase transition.
  We find that $\beta_{l}$, $\beta_{d}$, and $\beta_c$ change with the (mean) 
  vertex degree in a non-monotonic way; $\beta_d$ is
  distinct from $\beta_c$ for regular random graphs of vertex degrees
  $K\geq 64$, while $\beta_d$ are always identical to $\beta_c$ for
  Erd\"os-R\'enyi graphs (at least up to mean vertex degree $c=512$). 
  We also compute the minimum FVS size of regular random graphs through the
  zero-temperature first-step replica-symmetry-breaking mean field
  theory and reach good agreement with the results obtained on single graph
  instances by the belief propagation-guided decimation algorithm.
  Taking together, this paper presents a systematic theoretical study
  on the energy landscape property of a spin glass system with global cycle
  constraints.
\end{abstract}

\pacs{05.70.Fh, 75.10.Nr, 89.20.Ff}

\maketitle

\section{Introduction}

An undirected graph is formed by a set of
vertices and a set of undirected edges between pairs of vertices.
A cycle (or a loop) of such a graph is a closed path connected by
a set of different edges.
A feedback vertex set (FVS) is a subset of
vertices  intersecting with every cycle of this graph
\cite{Festa-Pardalos-Resende-1999}.
If all the vertices in a FVS are deleted from the graph,
there will be no cycle in the remaining subgraph.
The FVS problem
aims at constructing a FVS of cardinality (size)
not exceeding certain pre-specified value or proving the nonexistence of
it \cite{Festa-Pardalos-Resende-1999,NPC2}.
This problem has wide practical applications, such as combinatorial
circuit design \cite{Festa-Pardalos-Resende-1999}, deadlock
recovery in operating systems \cite{deadlock},
network dynamics analysis \cite{Fiedler-etal-2013,Mochizuki-etal-2013},
and epidemic spreading process \cite{FVSepidemic}.

The FVS problem is a combinatorial optimization problem in the
nondeterministic polynomial-complete (NP-complete) complexity class
\citep{NPC}.
It is generally believed (yet not rigorously proven) that there is no way to 
solve this problem by a complete algorithm in time bounded by a polynomial 
function of the number of vertices or edges in the graph. So far, the most
efficient complete algorithm is able to construct a FVS of global minimum
cardinality in an exponential time of the order $1.7548^{N}$
by solving the equivalent maximum induced forest problem \cite{exactlyB}.
Many heuristic algorithms have been developed to solve it approximately.
These algorithms are
incomplete in the sense that they may fail for some input graph instances, but
they have the merit of reaching a FVS solution very quickly if they succeed.
One famous heuristic algorithm is {\tt FEEDBACK} of Bafna and co-authors
\cite{FEEDBACK}, which is guaranteed, for any input graph, to construct a FVS
with cardinality at most two times the minimum value. In a more recent
paper, two of the present authors demonstrated that a heuristic algorithm
based on the idea of simulated annealing
extensively outperforms {\tt FEEDBACK} on random graphs and finite-dimensional
lattices \cite{EPJB2014}. The FVS problem has also been treated by statistical
physics methods and the associated belief propagation-guided decimation (BPD)
algorithm \cite{HJZEPJB2013}.
This physics-inspired message-passing algorithm outperforms simulated
annealing to some extent and constructs a FVS of cardinality being very
close to the global minimum value.

The spin glass model for the FVS problem was proposed
in \cite{HJZEPJB2013} by implementing the global cycle constraints as
a set of local edge constraints. This spin glass model was then studied by
mean field theory at the level of replica symmetry (RS)
without taking into account the possibility of spin glass phase transitions.
At low temperatures (equivalently, high
inverse temperatures $\beta$), the RS
mean field equations cannot reach a self-consistent solution
\cite{HJZEPJB2013}, which indicates  that the RS theory is valid only at
sufficiently high temperatures and that the property of the
FVS spin glass model at low temperature is much more complex than that at
high temperature.
In the present paper, we continue to study this spin glass
model at finite temperatures and at the zero temperature limit using
the first-step replica-symmetry-breaking (1RSB) mean field theory
\cite{PhysRevLett.75.2847,MEZ2001,Mezard-Montanari-2009}.

We mainly work on the ensemble of regular random (RR) graphs, and in some
cases also consider the ensemble of Erd\"os-R\'enyi (ER) random graphs.
The degree of a vertex is defined as the number of edges attached to the
vertex (i.e., the number of nearest neighbors).
Each vertex in a RR graph has the same degree $K$, but the edges in the
graph are connected completely at random. On the other hand, an ER
graph is created by setting up $M= (c/2) N$ edges completely at random
between $N$ vertices, where $c$ is the mean vertex degree.
When $N$ is sufficiently large, the degree of a randomly chosen vertex follows
the Poisson distribution with mean value $c$
\cite{Bollobas-2001}.

After reviewing the spin glass model and the RS mean field theory
in Sec.~\ref{sec:SGM}, we analyze the local stability of the RS theory
analytically (for RR graphs) and numerically (for ER graphs) in
Sec.~\ref{sec:stable} and investigate the dynamical (clustering) and static
(condensation) spin glass phase transitions in Sec.~\ref{sec:1RSB}. We determine
for each investigated graph ensemble the critical inverse temperature
$\beta_{l}$ for the local stability of
the RS theory, the inverse temperature $\beta_{d}$ of the clustering
(dynamical) transition,  and the inverse temperature $\beta_{c}$
of the condensation (static) transition. Both $\beta_{l}$ and $\beta_{d}$ change with the graph parameter $K$ or $c$
in a non-monotonic way. $\beta_{l}$ coincides with $\beta_d$
when $K$ or $c$ is relatively small; but $\beta_{l}$
exceeds $\beta_{d}$ when  $K\geq 40$ (for RR graphs) or
$c\geq 120$ (for ER graphs), which suggests that the RS mean field theory is
locally stable even after the system enters the spin glass phase.
For ER graph ensembles $\beta_{d}$ is indistinguishable from
$\beta_{c}$ for all the mean vertex degrees explored, while we notice that
$\beta_{c}$ becomes higher than $\beta_{d}$ for RR graph ensembles at
$K \geq 64$.
The existence of two distinct spin glass phases for dense RR graphs
may have significant algorithmic consequences.
In the final part of this paper we consider the $\beta \rightarrow \infty$ 
limit of the 1RSB mean field
theory (Sec.~\ref{sec:T01RSB}) to estimate the ensemble-averaged minimum
FVS cardinality. For RR graphs these theoretical results improve over the 
corresponding results obtained through the RS mean field theory, and they are 
in good agreement with numerical results obtained by the BPD algorithm.

Cycles are nonlocal properties of random graphs, and the cycle constraints in
the undirected FVS problem are therefore global in nature.
Phase transitions in globally constrained spin glass models and
combinatorial optimization problems are usually very hard to investigate.
We believe the results reported in this paper will also shed light on
the energy landscape properties of other globally constrained problems.

\section{Model and replica-symmetric theory}
\label{sec:SGM}

We consider an undirected simple graph $\mathcal{G}$ formed by $N$ vertices
and $M$ edges. There is no self-edge from a vertex to itself, and there is at
most one edge between any pair of vertices. For each vertex
$i\in \{1, 2, \ldots, N\}$ we denote by $\partial i$ the set of vertices that
are connected to $i$ through an edge. The degree $d_i$ of vertex $i$ is then
the cardinality of $\partial i$ ($d_i \equiv |\partial i|$).

\subsection{Local constraints and the partition function}

We assign a state $A_i$ to each vertex $i$ of the graph $\mathcal{G}$. $A_i$ can
take $(d_i+2)$ possible non-negative integer values from the union set
$\{0, i\}\cup \partial i$. If $A_i=0$ vertex $i$ is regarded as being empty,
otherwise it is occupied. In the latter case, if $A_i=i$ we say $i$ is a root
vertex, otherwise $A_i=j \in \partial i$ and we say $j$ is the parent vertex
of $i$ \cite{HJZEPJB2013}.

To represent the global cycle constraints in a distributed way, we define 
for each edge $(i, j)$ between vertices $i$ and $j$ a counting number
$C_{(i, j)}$ as
\begin{eqnarray}
  & & \hspace*{-0.8cm}
  C_{(i,j)}(A_i,A_j)   \equiv  \delta^0_{A_i}\delta^0_{A_j} \nonumber\\
  & &
  +\delta^0_{A_i}\bigl(1-\delta^0_{A_j}-\delta^i_{A_j}\bigr)
  +\delta^0_{A_j}\bigl(1-\delta^0_{A_i}-\delta^j_{A_i}\bigr) \nonumber  \\
  & &
  +\delta^j_{A_i} \bigl(1-\delta^0_{A_j}-\delta^i_{A_j}\bigr)
  + \delta^i_{A_j} \bigl(1-\delta^0_{A_i}-\delta^j_{A_i}\bigr) \; ,
  \label{eq:Cij}
\end{eqnarray}
where $\delta^a_b=1$ if $a=b$ and $\delta^a_b=0$ if $a\neq b$. This counting
number can take one of two possible values $0$ and $1$. We say that edge
$(i,j)$ is satisfied if and only if $C_{ij}(A_i,A_j)=1$, otherwise the edge
is regarded as unsatisfied \cite{HJZEPJB2013}. A microscopic configuration
$\underline{A}\equiv (A_1, A_2, \ldots,A_N)$ is called a legal
configuration if and only if it satisfies all the $M$ edges. A legal
configuration $\underline{A}$ has an important graphical property that each
connected component of the subgraph induced by all the occupied vertices of
$\underline{A}$ is either a tree (which has $n\geq 1$ vertices and $n-1$ edges)
or a so-called cycle-tree (which contains a single cycle and has $n\geq 2$
vertices and $n$ edges) \cite{HJZEPJB2013}.

Given a legal configuration $\underline{A}$ of graph $\mathcal{G}$, we can
easily construct a feedback vertex set $\Gamma$ as follows: (1) add
all the empty vertices of $\underline{A}$ to $\Gamma$; (2) if the subgraph
induced by the occupied vertices of $\underline{A}$ has one or more cycle-tree
components, then for each cycle-tree component we add a randomly chosen vertex
on the unique cycle to $\Gamma$ \cite{HJZEPJB2013}. On the other hand, given a
feedback vertex set $\Gamma$, we can easily construct many legal
configurations $\underline{A}$ as follows: (1) assign all the vertices
$i\in \Gamma$ the empty state $A_i=0$; (2) for each tree component (say $\tau$)
of the subgraph induced by all the vertices outside $\Gamma$, randomly choose
one vertex $j \in \tau$ as the root ($A_j=j$) and then determine the states of
all the other vertices in $\tau$ recursively: a nearest neighbor $k$ of $j$ has
state $A_k=j$ and a nearest neighbor $l \neq j$ of $k$ has state
$A_l=k$, and so on.

We define the energy of a microscopic configuration $\underline{A}$ as
\begin{equation}
  \label{eq:EnergyDef}
  E(\underline{A}) = \sum\limits_{i=1}^{N} \delta_{A_i}^{0} \; ,
\end{equation}
which just counts the total number of empty vertices. Because of the mapping
between legal configurations and feedback vertex sets, the energy function
$E(\underline{A})$ under the edge constraints (\ref{eq:Cij})  serves as a good 
proxy to the energy landscape of the undirected FVS problem.
The minimum value of $E(\underline{A})$ over all legal configurations 
is referred to as the ground-state (GS)
energy and is denoted as $E_0$. The corresponding configurations are the
GS configurations and the  number of all GS configurations is denoted
as $\Omega_0$. Due to the effect of cycle-trees the GS energy $E_0$ might be
slightly lower than the cardinality of a minimum FVS, but the
difference is negligible for $N$ sufficiently large \cite{HJZEPJB2013}.

The partition function of our spin glass model is
\begin{equation}
  \label{eq:PF}
  Z(\beta)=\sum_{\underline{A}} \exp\big[-\beta E(\underline{A})\big]
  \prod_{(i,j)\in \mathcal{G}}C_{(i,j)}(A_i,A_j) \; ,
\end{equation}
where $\beta$ is the inverse temperature. Notice that an illegal configurations
$\underline{A}$ contributes nothing to $Z(\beta)$, therefore $Z(\beta)$ is the 
sum of the statistical weights $e^{-\beta E(\underline{A})}$ of all legal 
configurations $\underline{A}$. The equilibrium probability of observing a
legal configuration $\underline{A}$ is then
\begin{equation}
  \mu(\underline{A}) = \frac{1}{Z} \exp\bigl[-\beta E(\underline{A}) \bigr]
  \prod_{(i,j)\in \mathcal{G}}C_{(i,j)}(A_i,A_j) \; .
\end{equation}
The total free energy of the system is related to the partition function through
\begin{equation}
  \label{eq:Fdef}
  F(\beta) = -\frac{1}{\beta}\ln Z(\beta) \; .
\end{equation}
The free energy has the limiting expression of
$F = E_0 - \frac{1}{\beta}\ln \Omega_0$ as $\beta$ approaches infinity.

\subsection{The belief-propagation equation}

The RS mean field theory assumes that all the equilibrium configurations of the
spin glass model (\ref{eq:PF}) form a single macroscopic state
\cite{Mezard-Montanari-2009}. The states of two or more distantly separated
vertices are then regarded as uncorrelated and their joint distribution is
expressed as the product of individual vertices' marginal distributions. Let us
denote by $q_i^{A_i}$ the marginal probability of vertex $i$'s state being $A_i$.
The state $A_i$ is of course strongly affected by the states of $i$'s nearest
neighbors, and the states of the vertices in $\partial i$ are also strongly
correlated since all of these vertices interact with $i$. Due to the local
tree-like structure of random graphs (i.e., cycle lengths are of order $\ln N$),
if vertex $i$ is removed, the vertices in set $\partial i$ will become distantly
separated and their states may then be assumed as uncorrelated.
For two vertices
$i$ and $j$ connected by an edge $(i, j)$,  let us denote by
$q_{j\rightarrow i}^{A_j}$ the marginal probability of $j$ being in state $A_j$ in
the absence of $i$ (this probability is referred to as a cavity probability).
After considering the interactions of $i$ with all the
vertices in $\partial i$, the RS theory then predicts that \cite{HJZEPJB2013}
\begin{subequations}
  \label{eq:cavity}
  \begin{align}
    q^0_i & =  \frac{e^{-\beta}}{z_i} \; , \\
    q^i_i & =  \frac{1}{z_i} \prod_{j \in \partial i} \bigl[
      q^0_{j\rightarrow i} +q^j_{j \rightarrow i}\bigr] \; , \\
    q^j_i & =  \frac{(1-q^0_{j\rightarrow i})}{z_i}
    \prod_{k \in \partial i\backslash j} \bigl[
      q^0_{k\rightarrow i}+q^k_{k \rightarrow i} \bigr]
    \; , \quad (j\in \partial i) 
  \end{align}
\end{subequations}
where $\partial i\backslash j$ means the set obtained by deleting vertex $j$
from the set $\partial i$, and the normalization factor $z_i$ is expressed as
\begin{equation}
  z_{i}\equiv e^{-\beta}+
  \Bigl[ 1+\sum_{k \in \partial i}
    \frac{1-q^0_{k\rightarrow i}}{q^0_{k\rightarrow i}+q^{k}_{k \rightarrow i}}
    \Bigr]
  \prod_{j \in \partial i}\bigl[q^0_{j\rightarrow i}+q^j_{j \rightarrow i}\bigr]
  \; .
\end{equation}
Similarly, the probabilities $q_{i\rightarrow j}^{A_i}$ and $q_{j\rightarrow i}^{A_j}$
on all the edges $(i, j)$ can be self-consistently determined by a set of
belief-propagation (BP) equations \cite{HJZEPJB2013}:
\begin{subequations}
  \label{eq:BP}
  \begin{align}
    q^0_{i\rightarrow j}& =
    \frac{e^{-\beta}}{z_{i\rightarrow j}} \; , \label{eq:BPe} \\
    q^i_{i\rightarrow j} & =
    \frac{1}{z_{i\rightarrow j}}
    \prod_{k \in \partial i\backslash j} \bigl[
      q^0_{k\rightarrow i}+q^k_{k \rightarrow i}\bigr]
    \; , \label{eq:BPr} \\
    q^l_{i\rightarrow j} & =  \frac{(1-q^0_{l\rightarrow i})}{z_{i\rightarrow j}}
    \prod_{k \in \partial i\backslash j,l} \bigl[
      q^0_{k\rightarrow i}+q^k_{k \rightarrow i}\bigr]
    \; , \quad  (l\in \partial i \backslash j)
    \label{eq:BPo}
  \end{align}
\end{subequations}
where $\partial i\backslash j, l$ means the set obtained by deleting vertices
$j$ and $l$ from  $\partial i$, and
\begin{equation}
  z_{i\rightarrow j}\equiv e^{-\beta}+ \Bigl[1+\sum_{k \in \partial i\backslash j}
    \frac{1-q^0_{k\rightarrow i}}{q^0_{k\rightarrow i}+q^k_{k \rightarrow i}} \Bigr]
  \prod_{l \in \partial i\backslash j}\bigl[ q^0_{l\rightarrow i}+q^l_{l \rightarrow i}
    \bigr] \; .
  \label{eq:zij}
\end{equation}
In our later discussions, Eq.~(\ref{eq:BP}) will be abbreviated as
$q_{i\rightarrow j} = b\big(\{q_{k\rightarrow i}: k\in \partial i\backslash j\} \bigr)$.

At a fixed point of the BP equation (\ref{eq:BP}) we can evaluate the total
free energy (\ref{eq:Fdef}) as the sum of contributions from all vertices
minus that from all edges \cite{HJZEPJB2013}:
\begin{equation}
  \label{eq:Frs}
  F = \sum\limits_{i=1}^{N} f_i - \sum\limits_{(i,j)\in \mathcal{G}} f_{i j} \; .
\end{equation}
The free energy contributions $f_i$ and $f_{i j}$ of a vertex $i$ and
an edge $(i, j)$ are computed through
\begin{eqnarray}
  \label{eq:RSFE}
  f_i & = & - \frac{1}{\beta} \ln \biggl[ e^{-\beta}+
    \Bigl( 1+\sum_{k \in \partial i}
    \frac{1-q^0_{k\rightarrow i}}{q^0_{k\rightarrow i}+q^{k}_{k \rightarrow i}}
    \Bigr) \nonumber \\
    & & \quad \quad\quad\quad\quad\quad \quad \quad \times
    \prod_{j \in \partial i}\bigl[q^0_{j\rightarrow i}+q^j_{j \rightarrow i}\bigr]
    \biggr] \; , \\
  f_{ij} & = & -\frac{1}{\beta}
  \ln \Bigl[ q^0_{i\rightarrow j} q^0_{j\rightarrow i}
    + (1-q^0_{i\rightarrow j})(q^0_{j\rightarrow i}+q^j_{j\rightarrow i})
    \nonumber \\
    & & \quad \quad \quad \quad
    +(1-q^0_{j\rightarrow i})(q^0_{i\rightarrow j}+q^i_{i\rightarrow j})
    \Bigr] \; .
\end{eqnarray}

The RS mean field equations (\ref{eq:cavity}), (\ref{eq:BP}) and (\ref{eq:Frs})
are applicable to single graph instances. We can also use these equations to
obtain the ensemble-averaged values for the free energy density, the mean 
energy density, and other thermodynamic quantities.
At the thermodynamic limit of $N\rightarrow \infty$ a random graph is
characterized by the vertex degree distribution $P(d)$, which is the 
probability of a randomly chosen vertex having $d$ nearest neighbors.
As there is no degree correlation in a random graph, the probability $Q(d)$
that the degree of the vertex at the end of a randomly chosen edge being $d$
is related to $P(d)$ through
\begin{equation}
  Q(d) = \frac{d \, P(d)}{\sum_{d^\prime \geq 1} d^\prime \, P(d^\prime)}
    \quad \quad \quad (d\geq 1) \; .
\end{equation}
For the RR ensemble $P(d) = Q(d)= \delta_{d}^{K}$;
for the ER ensemble $P(d)= \frac{e^{-c} c^d}{d!}$ and 
$Q(d) = \frac{e^{-c} c^{d-1}}{(d-1)!}$ (both are Poisson distributions).

Let us denote by $\mathcal{P}[q_{i\rightarrow j}]$ the probability functional
of the cavity probability function $q_{i\rightarrow j}$, which gives the
probability that a randomly picked edge $(i, j)$ of the graph has the
cavity probability function $q_{i\rightarrow j}$. This probability functional is
governed by the following self-consistent equation:
\begin{equation}
  \mathcal{P}[q] = \sum\limits_{d=1}Q(d) \int \prod\limits_{j=1}^{d-1}
  \mathcal{D}q_{j} \mathcal{P}[q_j]
  \delta\bigl[ q-b(\{q_{j}\})\bigr] \; ,
\end{equation}
where the Dirac delta functional $\delta\bigl[ q - b(\{q_j\}) \bigr]$ 
ensures that the output cavity probability function $q$ and the set of input 
cavity  probability functions $\{q_j\}$ are related by the BP equation 
(\ref{eq:BP}). For the RR ensemble, $\mathcal{P}[q]$ is simply a Dirac delta 
functional.

\section{Local stability of the replica-symmetric theory}
\label{sec:stable}

Before studying the undirected FVS problem by the 1RSB mean field theory, 
let us first check the local stability of the RS mean field theory.
Assume that a fixed point solution, say
$\{\tilde{q}_{i\rightarrow j},\, \tilde{q}_{j\rightarrow i}\, :\,
(i, j)\in \mathcal{G}\}$,
of the BP equation (\ref{eq:BP}) has been reached. We perform a
perturbation to this fixed point, for example,
\begin{equation}
  \label{eq:eps}
  q_{i\rightarrow j}^0 = \tilde{q}_{i\rightarrow j}^0 +
  \epsilon_{i \rightarrow j}^0 \; , \quad \quad
  q_{i\rightarrow j}^i = \tilde{q}_{i\rightarrow j}^i +
  \epsilon_{i\rightarrow j}^i \; ,
\end{equation}
with $\epsilon_{i\rightarrow j}^0$ and $\epsilon_{i\rightarrow j}^i$ being
sufficiently small. If the magnitudes of all these small quantities
$\epsilon_{i\rightarrow j}^0$ and $\epsilon_{i\rightarrow j}^i$ shrink during the
iteration of Eqs.~(\ref{eq:BPe}) and (\ref{eq:BPr}),  $q_{i\rightarrow j}^0$
and $q_{i\rightarrow j}^i$ will relax back to $\tilde{q}_{i\rightarrow j}^0$ and
$\tilde{q}_{i\rightarrow j}^i$, and $q_{i\rightarrow j}^l$ for
$l\in \partial i\backslash j$ will relax back to $\tilde{q}_{i\rightarrow j}^l$
following Eq.~(\ref{eq:BPo}). If such a converging situation occurs, we 
say that the BP fixed point is locally stable under perturbations, otherwise
it is locally unstable
\cite{montanari2004,montanari2003,Lenka2006,LenkaThe,PhysRevE.80.021122}.

\subsection{Regular random graphs}

For RR graphs the BP fixed point is easy to determine, namely 
all $\tilde{q}_{i\rightarrow j}^0 = \tilde{q}^{O}$ and
$\tilde{q}_{i\rightarrow j}^i = \tilde{q}^{R}$, with
\begin{subequations}
  \begin{align}
    \tilde{q}^{O} & = \frac{e^{-\beta}}
          {e^{-\beta}+(\tilde{q}^O+\tilde{q}^R)^{K-2} \bigl[
              K-1+\tilde{q}^R - (K-2) \tilde{q}^O\bigr]}\; , \\
          \tilde{q}^{R} & = \frac{(\tilde{q}^{O}+\tilde{q}^{R})^{K-1}}
                {e^{-\beta}+(\tilde{q}^O+\tilde{q}^R)^{K-2} \bigl[
                    K-1+\tilde{q}^R - (K-2) \tilde{q}^O\bigr]} \; .
  \end{align}
\end{subequations}
We can combine Eq.~(\ref{eq:eps}) with Eq.~(\ref{eq:BP})
and write an iterative equation of perturbation as
\begin{equation}
  \label{eq:stab1}
  \begin{bmatrix}
    \epsilon^0_{i\rightarrow j}\\
    \\
    \epsilon^i_{i\rightarrow j}
  \end{bmatrix}
  = \textbf{J} \times
  \begin{bmatrix}
    \sum\limits_{k\in \partial i \setminus j} \epsilon^0_{k\rightarrow i}\\
    \\
    \sum\limits_{k\in \partial i \setminus j} \epsilon^k_{k\rightarrow i}
  \end{bmatrix},
\end{equation}
where
\begin{equation}
  \textbf{J}=
  \begin{bmatrix}
    \frac{\partial q_{i\rightarrow j}^0}{\partial q_{k\rightarrow i}^0 }
    & \frac{\partial q_{i\rightarrow j}^0}{\partial q_{k\rightarrow i}^k} \\
    & \\
    \frac{\partial q_{i\rightarrow j}^i}{\partial q_{k\rightarrow i}^0}
    & \frac{\partial q_{i\rightarrow j}^i}{\partial q_{k\rightarrow i}^k}
  \end{bmatrix}
\end{equation}
is a $2\times 2$ matrix evaluated at the BP fixed point, whose largest absolute
eigenvalue of matrix $\textbf{J}$ is denoted as $\lambda$.
It is reasonable to assume that the perturbations 
$\epsilon_{i\rightarrow j}^0$ and $\epsilon_{i\rightarrow j}^i$ follow the 
distributions with mean value $0$  and variance $\sigma_O^2$ and
$\sigma_R^2$ respectively. Then the mean values of
$\sum\limits_{k\in \partial i \setminus j} \epsilon^0_{k\rightarrow i}$ and
$\sum\limits_{k\in \partial i \setminus j} \epsilon^R_{k\rightarrow i}$ are 
still $0$, and their variances are $(K-1)\sigma_O^2$ and
$(K-1)\sigma_R^2$ respectively. After one iteration with Eq.~(\ref{eq:stab1})
the variance of $\epsilon_{i\rightarrow j}^0$ and
$\epsilon_{i\rightarrow j}^i$ must not exceed 
$\lambda^2(K-1)\sigma_O^2$ and $\lambda^2(K-1)\sigma_R^2$. 
Considering that the perturbation should shrink to $0$ in the case of 
stability, we have the local stability criterion that $(K-1)\lambda^2<1$.

\subsection{Erd\"os-R\'enyi random graphs}

Because the vertex degree dispersion in the ER graph ensemble, the analytical
method we used in the preceding subsection is not applicable here.
Therefore, we can only measure the magnitude of the perturbations during 
the BP iteration and then figure out the region where the BP 
equation (\ref{eq:BP}) is stable (see Appendix E~1 of \cite{LenkaThe}).
This numerical procedure starts from running population dynamics for BP a 
sufficiently long time to reach a steady state. 
After that, we make a replica of the whole population to get two identical 
populations and then perturb one of the populations slightly. Finally
we continue to perform BP population dynamics simulations
starting from these two initial populations using the same sequence of 
random numbers. If the two populations finally converge to each other during 
the iteration process we say that the BP equation is locally stable, 
otherwise it is regarded as locally unstable.

For RR graph ensemble we have checked that the results obtained by such a
stability analysis are identical to those obtained by the criterion of 
the preceding subsection.

\begin{figure}
\includegraphics[width=0.45\textwidth]{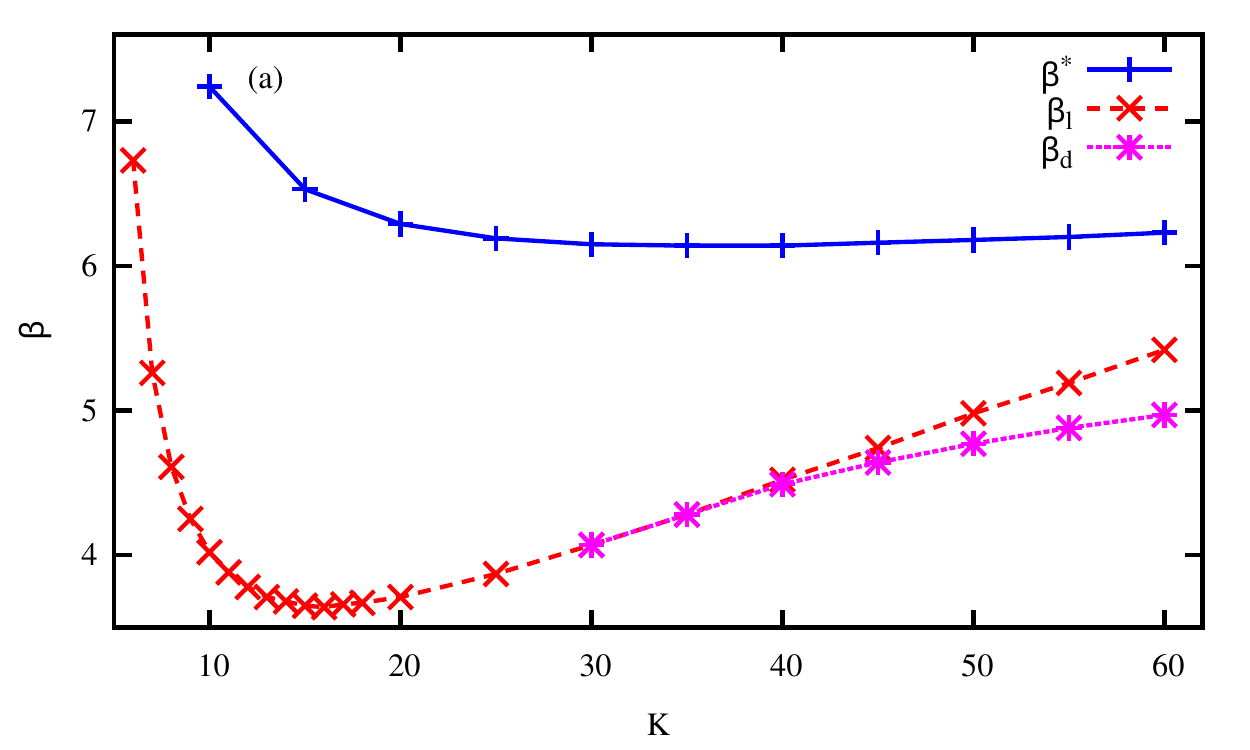}
\includegraphics[width=0.45\textwidth]{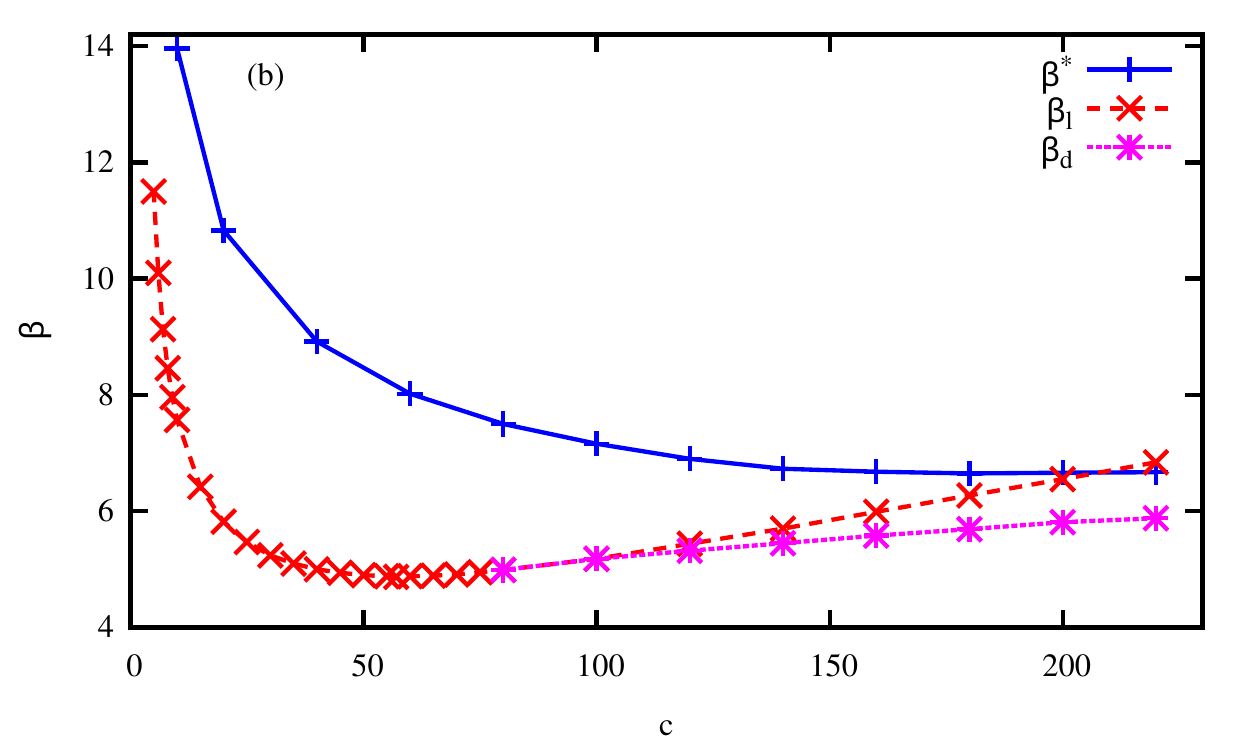}
\caption{
\label{fig:PTpoint}
The local stability critical inverse temperature $\beta_{l}$ of the
RS mean field theory, the dynamical transition inverse temperature $\beta_d$
and the inverse temperature $\beta^*$ where the RS entropy equals
to $0$. (a) Regular random graph ensemble with integer vertex degree $K$;
(b) Erd\"os-R\'enyi graph ensemble with mean vertex degree $c$.
}
\end{figure}

\subsection{Local stability results}

In Fig.~\ref{fig:PTpoint}, we compare the local stability critical
inverse temperature $\beta_{l}$ with the dynamical transition
inverse temperature $\beta_d$ (obtained in Sec.~\ref{sec:1RSB})
and the inverse temperature $\beta^*$ where the RS entropy density
equals to zero.
We find that $\beta_{l}$ is not a monotonic function of the degree $K$ (for
RR graphs) or the mean degree $c$ (for ER graphs), but it first decreases with
$K$ or $c$ and then slowly increases as
$K$ goes beyond $16$ or $c$ goes beyond $57$.
The dynamical transition point $\beta_d$ coincides with $\beta_{l}$ 
when $K<40$ (for RR graphs) or $c<120$ (for ER graphs).
When $K$ or $c$ becomes larger we find that $\beta_l > \beta_d$, that is,
the RS mean field solution is still locally stable as the system enters the
spin glass phase.

\section{The dynamical transition and condensation transition}
\label{sec:1RSB}

We now investigate spin glass phase transitions in the undirected FVS
model (\ref{eq:PF}). For simplicity of numerical computations we restrict 
our discussion to the ensemble of RR graphs with integer vertex degrees $K$.

\subsection{The first-step replica-symmetry-breaking mean field theory}

We first give a brief review of the 1RSB mean field theory of spin glasses
\cite{Mezard-Montanari-2009}. According to this theory, at sufficiently high
inverse temperatures the space of legal configurations of the energy function
(\ref{eq:EnergyDef}) may break into exponentially many subspaces, each of which
corresponds to a macroscopic state of the system and contains a set of
relatively similar legal configurations. In this subsection, unless otherwise
stated, we discuss the system at the level of macroscopic states. The partition
function of macroscopic state $\alpha$ is defined as
\begin{equation}
  \label{eq:Zbeta1}
  Z_\alpha(\beta)=e^{-\beta F_\alpha}
  =\sum\limits_{\underline{A}\in\alpha}
  \exp\big[-\beta E(\underline{A})\big]  \; ,
\end{equation}
where the sum runs over all the legal configurations in the macroscopic
state $\alpha$, and $F_\alpha$ is the free energy of $\alpha$
\cite{Mezard-Montanari-2009,MEZ2001}.

We can define a Boltzmann distribution at the level of macroscopic
states as
\begin{equation}
  \omega_\alpha \equiv \frac{ e^{- y F_\alpha}}{\Xi(y; \beta)} \; .
  \label{eq:omegaXi}
\end{equation}
The parameter $y$ is the inverse temperature at the macroscopic level,
which may be different from the inverse temperature $\beta$ at the
level of microscopic configurations. The ratio between $y$ and $\beta$, namely
$m \equiv \frac{y}{\beta}$ is referred to as the Parisi parameter
\cite{Mezard-Montanari-2009}. The quantity $\omega_{\alpha}$ as defined in
Eq.~(\ref{eq:omegaXi}) determines the weight of macroscopic state $\alpha$ among
all the macroscopic states, and the normalization constant 
$\Xi(y; \beta)\equiv \sum\limits_{\alpha} e^{- y F_\alpha}$ is
the partition function the level of macroscopic states,
which can also be calculated by the integration
\begin{equation}
  \Xi(y; \beta)=\int df e^{N[-yf+\Sigma(f)]} \; ,
\end{equation}
where $f$ is the free energy density of a macroscopic state, and $\Sigma(f)$,
called the complexity, is the entropy density of macroscopic states with
free energy density $f$. The behavior of complexity plays an important role in 
determining whether the system has a dynamical phase transition or 
not \cite{PhysRevB.36.5388}.
We can define the grand free energy $G(y; \beta)$ of the system as
\begin{equation}
  G(y; \beta) \equiv  - \frac{1}{y} \ln \Xi(y; \beta) \; .
\end{equation}

The mean free energy $\langle F \rangle$ is the mean free energy of 
macroscopic states according to the distribution (\ref{eq:omegaXi}).
In the thermodynamic limit $N\rightarrow \infty$, the macroscopic states
with the free energy density 
$f = \argmax_f \bigl[-y f+\Sigma(f)\bigr]$ dominate the partition function
$\Xi(y; \beta)$, and then $G(y; \beta)=N \Bigl[f-\Sigma(f)/y\bigr]$ and
$\langle F \rangle=N f$. Therefore the complexity is obtained through
\begin{equation}
  \Sigma =\frac{y }{N} \Bigl[ \langle F \rangle - G \Bigr] \; .
\end{equation}

When there are many macroscopic states, for a given edge $(i,j)$ the
cavity message $q_{i\rightarrow j}$ from vertex $i$ to vertex $j$ may be 
different in different macroscopic states. The distribution of this message 
among all the macroscopic states is denoted as
$Q_{i\rightarrow j}[q_{i\rightarrow j}]$. Under the distribution of
Eq.~(\ref{eq:omegaXi}) and for random graphs, we have the following
self-consistent equation for $Q_{i\rightarrow j}[q_{i\rightarrow j}]$, which is
referred to as the survey propagation (SP) equation:
\begin{eqnarray}
  Q_{i\rightarrow j}[q_{i\rightarrow j}] & = &
  \frac{1}{\Xi_{i\rightarrow j}}
  \int \prod\limits_{k\in \partial i\backslash j} \mathcal{D}
       {q}_{k \rightarrow i}\, Q_{k \rightarrow i}[q_{k\rightarrow i}]\,
       e^{-y f_{i\rightarrow j}}\nonumber \\
       & & \times\, \delta\Bigl[q_{i\rightarrow j}-
         b\bigl(\{{q}_{k\rightarrow i}: k\in \partial i\backslash j\}
         \bigr) \Bigr] \; ,
       \label{eq:1RSB}
\end{eqnarray}
where $f_{i\rightarrow j}$ is the free energy change associated with the
interactions of vertex $i$ with the vertices in the set
$\partial i\backslash j$,
\begin{eqnarray}
  f_{i\rightarrow j} & = & - \frac{1}{\beta} \ln z_{i\rightarrow j}
\end{eqnarray}
with $z_{i\rightarrow j}$ computed through Eq.~(\ref{eq:zij}); and the
normalization constant $\Xi_{i\rightarrow j}$ is determined through
\begin{equation}
  \Xi_{i\rightarrow j} = \int \prod\limits_{k\in \partial i\backslash j}
  \mathcal{D} q_{k \rightarrow i}\,
  Q_{k \rightarrow i}[q_{k\rightarrow i}]\,
  e^{-y f_{i\rightarrow j}} \; .
\end{equation}

After a fixed-point solution of Eq.~(\ref{eq:1RSB}) is obtained, the grand
free energy $G$ and the mean free energy $\langle F \rangle$
can be computed, respectively, through
\begin{eqnarray}
  G &=& \sum\limits_{i=1}^{N} g_i-\sum\limits_{(i, j) \in \mathcal{G}} g_{i j} \; ,
  \\
  \langle F \rangle &=&
  \sum\limits_{i=1}^{N} \langle f_i\rangle
  - \sum\limits_{(i, j) \in \mathcal{G}} \langle f_{i j}\rangle \; .
\end{eqnarray}
In these  equations $g_i$ and $\langle f_i \rangle$ are the grand free
energy and mean free energy contribution from a vertex $i$, while
$g_{i j}$ and $\langle f_{i j} \rangle$ are the corresponding contributions
from an edge $(i, j)$. The explicit expressions of these quantities
read
\begin{eqnarray}
  g_i &=& -\frac{1}{y}\log\bigg[\int \prod_{j\in\partial i}
    \mathcal{D} q_{j\rightarrow i} Q_{j\rightarrow i}[q_{j\rightarrow i}]
    e^{-yf_i}\bigg] \; , \\
  g_{ij} &=& -\frac{1}{y}\log\bigg[\int \mathcal{D} q_{j\rightarrow i}\nonumber
    \mathcal{D} q_{i\rightarrow j} Q_{j\rightarrow i}[q_{j\rightarrow i}]\\
    & &\quad \quad \quad  \quad \quad \quad  
    \times Q_{i\rightarrow j}[q_{i\rightarrow j}] e^{-yf_{ij}}\bigg] \; , \\
  \langle f_i \rangle &=&
  \frac{\int \prod_{j\in\partial i}\mathcal{D} q_{j\rightarrow i}
    Q_{j\rightarrow i}[q_{j\rightarrow i}] f_i e^{-yf_i}}{\int
    \prod_{j\in\partial i}\mathcal{D} q_{j\rightarrow i}
    Q_{j\rightarrow i}[q_{j\rightarrow i}] e^{-yf_i}} \; , \\
  \langle f_{i j} \rangle &=&\frac{\int \mathcal{D} q_{j\rightarrow i} 
    \mathcal{D} q_{i\rightarrow j} Q_{j\rightarrow i}[q_{j\rightarrow i}]
    Q_{i\rightarrow j}[q_{i\rightarrow j}] f_{i j} e^{-yf_{i j}}}{\int 
    \mathcal{D} q_{j\rightarrow i} \mathcal{D} q_{i\rightarrow j}
    Q_{j\rightarrow i}[q_{j\rightarrow i}] Q_{i\rightarrow j}[q_{i\rightarrow j}]
    e^{-yf_{i j}}} \; . \nonumber\\ \;
\end{eqnarray}

\subsection{The special case of $y = \beta$}

We now consider the most natural value of $y=\beta$ (the inverse temperatures
at the level of macroscopic states and at the level of microscopic
configurations are exactly equal) and investigate spin glass phase transitions.
In order to simplify the derivation of the 1RSB mean field theory at 
$y=\beta$, we introduce a coarse-grained probability as
\begin{equation}
  \label{eq:defX}
  q_{i\rightarrow j}^X \equiv 
  \sum_{l\in \partial i\setminus j} q_{i\rightarrow j}^l \; ,
\end{equation}
where the superscript `X' means that the state $A_i$ of vertex $i$ is 
neither $i$ nor $0$.
The quantity $q_{i\rightarrow j}^X$ gives the probability that, in the
absence of vertex $j$,  vertex $i$ is occupied ($A_i > 0$)
but is not a root ($A_i \neq i$).

Following the work of M\'ezard and Montanari \cite{Mezard2006} we define
$\overline{q}_{i\rightarrow j} \equiv \bigl(\overline{q}_{i\rightarrow j}^{0},
\overline{q}_{i\rightarrow j}^{i}, \overline{q}_{i\rightarrow j}^{X} \bigr)$
as the mean value of the probability $q_{i\rightarrow j}$ among all the
macroscopic states, with
\begin{subequations}
  \label{eq:averageqij}
  \begin{align}
    \overline{q}_{i\rightarrow j}^0  & =
    \int \mathcal{D} q_{i\rightarrow j}
    \, Q_{i\rightarrow j}[q_{i\rightarrow j}] \, q_{i\rightarrow j}^0 \; , \\
    \overline{q}_{i\rightarrow j}^i  & = \int \mathcal{D} q_{i\rightarrow j}
    \, Q_{i\rightarrow j}[q_{i\rightarrow j}]\, q_{i\rightarrow j}^i \; , \\
    \overline{q}_{i\rightarrow j}^X  & = \int \mathcal{D} q_{i\rightarrow j}
    \, Q_{i\rightarrow j}[q_{i\rightarrow j}]\, q_{i\rightarrow j}^X \; .
  \end{align}
\end{subequations}
$\overline{q}_{i\rightarrow j}^0$ and $\overline{q}_{i\rightarrow j}^i$ are the
mean probabilities of vertex $i$ being in state $A_i=0$ and $A_i=i$,
respectively, in the absence of vertex $j$, while
$\overline{q}_{i\rightarrow j}^X$ is this vertex's mean probability of taking
states different from $A_i=0$ and $A_i=i$ in the absence of vertex $j$. In
addition, we define three auxiliary conditional probability functionals as
\begin{subequations}
  \label{eq:Qijcond}
  \begin{align}
    Q_{i\rightarrow j}^0 \bigl[q_{i\rightarrow j} \bigr] & \equiv
    \frac{ q_{i\rightarrow j}^0 \, Q_{i\rightarrow j}\bigl[q_{i\rightarrow j}\bigr] }
         {\overline{q}_{i\rightarrow j}^{0}} \; ,
         \label{eq:Qijcond0} \\
         Q_{i\rightarrow j}^i\bigl[ q_{i\rightarrow j} \bigr] & \equiv
         \frac{q_{i\rightarrow j}^i\, Q_{i\rightarrow j}\bigl[ q_{i\rightarrow j}
             \bigr]}{\overline{q}_{i\rightarrow j}^i} \; ,
         \label{eq:Qijcondi} \\
         Q_{i\rightarrow j}^X\bigl[ q_{i\rightarrow j} \bigr] & \equiv
         \frac{ q_{i\rightarrow j}^X\, Q_{i\rightarrow j} \bigl[ q_{i\rightarrow j}
             \bigr]}{\overline{q}_{i\rightarrow j}^X } \; .
         \label{eq:Qijcondx}
  \end{align}
\end{subequations}
$Q_{i\rightarrow j}^{0}\bigl[ q_{i\rightarrow j} \bigr]$, 
$Q_{i\rightarrow j}^{i}\big[ q_{i\rightarrow j} \bigr]$  and
$Q_{i\rightarrow j}^X\bigl[q_{i\rightarrow j} \bigr]$ are the distribution 
functionals for the cavity message $q_{i\rightarrow j}$ under 
the condition of $A_i=0$, $A_i=i$ and $A_i \notin \{0, i\}$, respectively.
It is easy to verify the identity that
\begin{eqnarray}
  Q_{i\rightarrow j}\bigl[ q_{i\rightarrow j} \bigr] & = &
  \overline{q}_{i\rightarrow j}^0
  \, Q_{i\rightarrow j}^0 \bigl[ q_{i\rightarrow j} \bigr] +
  \overline{q}_{i\rightarrow j}^i
  \, Q_{i\rightarrow j}^i \bigl[ q_{i\rightarrow j} \bigr] \nonumber \\
  & & \quad \quad + \overline{q}_{i\rightarrow j}^X
  \, Q_{i\rightarrow j}^X \bigl[ q_{i\rightarrow j} \bigr] \; .
\end{eqnarray}

At the special case of $y=\beta$,  by inserting the SP
equation (\ref{eq:1RSB}) into Eq.~(\ref{eq:averageqij}), we obtain that the
mean cavity probability $\overline{q}_{i\rightarrow j}$ also obeys the BP
equation (\ref{eq:BP}),
\begin{equation}
  \label{eq:Qidentity}
  \overline{q}_{i\rightarrow j} = b\bigl(\{\overline{q}_{k\rightarrow i}
  : k\in \partial i\backslash j\}\bigr) \; .
\end{equation}
In other words, the mean cavity probabilities
$\{\overline{q}_{i\rightarrow j}\}$ can be computed without the need of
 computing the probability functionals
$\{Q_{i\rightarrow j}[q_{i\rightarrow j}]\}$.
We now exploit this nice property in combination with Eq.~(\ref{eq:Qijcond})
to greatly simplify the numerical difficulty of implementing the 1RSB mean
field theory \cite{Mezard2006,Florent2007}.

After inserting Eq.~(\ref{eq:1RSB}) into Eq.~(\ref{eq:Qijcond}), we find that
the three auxiliary probability functionals obey the following self-consistent
equations:
\begin{subequations}
  \label{eq:QcondBP}
  \begin{align}
    &  Q_{i\rightarrow j}^0[q_{i\rightarrow j}]   =
    \prod\limits_{k \in \partial i \backslash j}
    \int \mathcal{D} q_{k\rightarrow i}\,
    \Bigl[\overline{q}_{k\rightarrow i}^0 \, Q_{k\rightarrow i}^0[
        q_{k\rightarrow i}]  \nonumber \\
      & \quad   + \overline{q}_{k\rightarrow i}^k \, Q_{k\rightarrow i}^k[
        q_{k\rightarrow i}] + \overline{q}_{k\rightarrow i}^X \,
      Q_{k\rightarrow i}^X [ q_{k\rightarrow i}] \Bigr]
    \nonumber \\
    & \quad \times  \delta\Bigl[ q_{i\rightarrow j} -
      b\bigl(\{q_{k\rightarrow i}\}\bigr) \Bigr] \; ,
    \label{eq:QcondBP0}  \\
    & Q_{i\rightarrow j}^i[ q_{i\rightarrow j}] =  \prod_{k\in
      \partial i \backslash j} \int \mathcal{D}{\bf q}_{k\rightarrow i}
    \Bigl[
      \frac{\overline{q}_{k\rightarrow i}^0} {\overline{q}_{k \rightarrow i}^0+
        \overline{q}_{k \rightarrow i}^k}
      Q_{k\rightarrow i}^0[q_{k\rightarrow i}] \nonumber \\
      & \quad + \frac{\overline{q}_{k\rightarrow i}^k}{
        \overline{q}_{k \rightarrow i}^0+ \overline{q}_{k \rightarrow i}^k}
      Q_{k\rightarrow i}^k[ q_{k\rightarrow i}] \Bigr]
    \ \delta\Bigl[ q_{i\rightarrow j} -
      b\bigl(\{{\bf q}_{k\rightarrow i}\} \bigr) \Bigr] \; ,
    \label{eq:QcondBPi}  \\
    & Q_{i\rightarrow j}^X[q_{i\rightarrow j}] =
      \sum_{k\in\partial i\backslash j}\omega_{k\rightarrow i} \int\mathcal{D}
      q_{k\rightarrow i}
      \Bigl[ \frac{\overline{q}_{k\rightarrow i}^k}{
          \overline{q}_{k \rightarrow i}^k + \overline{q}_{k \rightarrow i}^X}
        \nonumber \\
        & \quad \times
        Q_{k\rightarrow i}^k[q_{k\rightarrow i}]
        + \frac{\overline{q}_{k\rightarrow i}^X}{
          \overline{q}_{k \rightarrow i}^k + \overline{q}_{k \rightarrow i}^X}
        Q_{k\rightarrow i}^X[ q_{k\rightarrow i} ] \Bigr]
      \nonumber \\
      &\quad  \times
      \prod\limits_{l\in \partial i\backslash j, k}\int \mathcal{D}
      q_{l\rightarrow i} \Bigl[ \frac{
          \overline{q}_{l\rightarrow i}^0}
        {\overline{q}_{l \rightarrow i}^0 +
          \overline{q}_{l \rightarrow i}^l}
        Q_{l\rightarrow i}^0[q_{l\rightarrow i}] \nonumber \\
        & \quad+
        \frac{\overline{q}_{l\rightarrow i}^l}{
          \overline{q}_{l \rightarrow i}^0 +
                  \overline{q}_{l \rightarrow i}^l}
        Q_{l\rightarrow i}^l[q_{l\rightarrow i}] \Bigr]
      \  \delta\Bigl[ q_{i\rightarrow j} -
        b\bigl(\{q_{k\rightarrow i}\} \bigr) \Bigr] \; .
      \label{eq:QcondBPx}
  \end{align}
\end{subequations}
In Eq.~(\ref{eq:QcondBPx}) the probability $w_{k\rightarrow i}$ is
determined as
\begin{equation}
  \omega_{k\rightarrow i}
  =\frac{(1-\overline{q}_{k\rightarrow i}^0)\prod\limits_{l\in\partial i
      \backslash j,k}
    [\overline{q}_{l\rightarrow i}^0+\overline{q}_{l\rightarrow i}^l]}{
    \sum\limits_{m\in\partial i\backslash j}(1-\overline{q}_{m\rightarrow i}^0)
    \prod\limits_{l\in\partial i \backslash  j,m}[\overline{q}_{l\rightarrow i}^0
      + \overline{q}_{l\rightarrow i}^0]} \; ,
\end{equation}
and it can be understood as the probability of choosing 
vertex $k$ among all the vertices in the set $\partial i\backslash j$.
The iterative equation (\ref{eq:QcondBP}) avoids the difficulty of reweighted
sampling in the original SP equation (\ref{eq:1RSB}).

For a given graph instance $\mathcal{G}$,
we describe the statistical property of vertex $i$ in the absence of the
neighboring vertex $j$ by the mean cavity probability function
$\overline{q}_{i\rightarrow j}$ and the three conditional probability
functionals $Q_{i\rightarrow j}^{0}\bigl[ q\bigr]$,
$Q_{i\rightarrow j}^{i}\bigl[ q\bigr]$, and $Q_{i\rightarrow j}^{X}\bigl[ q \bigr]$,
each of which is represented by a set of sampled cavity probabilities
$q_{i\rightarrow j}$. We first iterate the BP equation (\ref{eq:Qidentity})
a number of rounds to bring the set of mean cavity probabilities
$\{\overline{q}_{i\rightarrow j}\}$ to the fixed point (or at least close to
the fixed point). Then Eq.~(\ref{eq:QcondBP}) is iterated to drive all the
conditional probability functionals to their steady states.
For example, to update $Q_{i\rightarrow j}^X\bigl[ q_{i\rightarrow j} \bigr]$ using
Eq.~(\ref{eq:QcondBPx}), we (1) choose a vertex $k \in \partial i \backslash j$
with probability $\omega_{k\rightarrow i}$, and then (2) draw a cavity
probability $q_{k\rightarrow i}$ from $Q_{k \rightarrow i}^{k}[q]$ with probability
$\frac{\overline{q}_{k\rightarrow i}^k}{\overline{q}_{k\rightarrow i}^k + 
  \overline{q}_{k\rightarrow i}^X}$
or from $Q_{k\rightarrow i}^X[q]$ with the remaining probability
$\frac{\overline{q}_{k\rightarrow i}^X}{\overline{q}_{k\rightarrow i}^k +
\overline{q}_{k\rightarrow i}^X}$,
and (3) for each of the other vertices
$l\in \partial i\backslash j, k$ we select a cavity probability
$q_{l\rightarrow i}$  from $Q_{l\rightarrow i}^{0}[q_{l\rightarrow i}]$ with 
probability
$\frac{\overline{q}_{l\rightarrow i}^0}{\overline{q}_{l\rightarrow i}^0 +
  \overline{q}_{l\rightarrow i}^i}$
or from $Q_{l\rightarrow i}^l[q_{l\rightarrow i}]$ with the remaining probability
$\frac{\overline{q}_{l\rightarrow i}^l}{\overline{q}_{l\rightarrow i}^0 +
\overline{q}_{l\rightarrow i}^l}$,
and finally (4) we generate a new cavity
probability $q_{i\rightarrow j}$ using the BP equation (\ref{eq:BP})
and replace a randomly chosen old cavity probability of the set
representing $Q_{i\rightarrow j}^X[q_{i\rightarrow j}]$ by this new one.
The other two conditional probability functionals 
$Q_{i\rightarrow j}^0 [q_{i\rightarrow j}]$ and
$Q_{i\rightarrow j}^i [q_{i\rightarrow j}]$ are updated following the same numerical
procedure.

At $y = \beta$ the computation of the grand free energy density
$g \equiv G/N$ and the mean free energy density
$\langle f \rangle \equiv \langle F \rangle / N$
can also be carried out without reweighting among the
different macroscopic states.
We list in Appendix~\ref{sec:appyb1} the explicit mean field expressions for
computing $g$ and $\langle f \rangle$.

The initial condition for the iterative equation (\ref{eq:QcondBP}) is chosen
to be the following set of $\delta$-formed probability functionals:
\begin{subequations}
  \label{eq:QcondInitial}
  \begin{align}
    Q_{i\rightarrow j}^0 [q_{i\rightarrow j}] & =
    \delta(q_{i\rightarrow j}^0 - 1)\,
    \delta(q_{i\rightarrow j}^i)\,
    \delta(q_{i\rightarrow j}^X)
    \; , \\
    Q_{i\rightarrow j}^i [q_{i\rightarrow j}] & =
    \delta(q_{i\rightarrow j}^0) \,
    \delta(q_{i\rightarrow j}^i - 1) \,
    \delta(q_{i\rightarrow j}^X)
    \; , \\
    Q_{i\rightarrow j}^X [q_{i\rightarrow j}] & =
    \delta(q_{i\rightarrow j}^0) \,
    \delta(q_{i\rightarrow j}^i) \,
    \delta(q_{i\rightarrow j}^X - 1)
    \; .
  \end{align}
\end{subequations}
According to the theoretical analysis in \cite{Mezard2006}, if the 
conditional probability functionals  (\ref{eq:Qijcond}) starting from this
initial condition converge to the trivial fixed point
\begin{eqnarray}
  &  \hspace*{-1.5cm} Q_{i\rightarrow j}^0 [q_{i\rightarrow j}]  =
  Q_{i\rightarrow j}^i [q_{i\rightarrow j}]  =
  Q_{i\rightarrow j}^X [q_{i\rightarrow j}]  = \nonumber \\
  &  \delta(q_{i\rightarrow j}^0 - \overline{q}_{i\rightarrow j}^0 ) \,
  \delta(q_{i\rightarrow j}^i - \overline{q}_{i\rightarrow j}^i ) \,
  \delta(q_{i\rightarrow j}^X-\overline{q}_{i\rightarrow j}^X) \; ,
  \label{eq:Qtrivial}
\end{eqnarray}
the system is then in the ergodic phase with a unique equilibrium 
macroscopic states (complexity $\Sigma = 0$). If the conditional probability
functionals  (\ref{eq:Qijcond})
converges to a fixed point different from Eq.~(\ref{eq:Qtrivial}), the
system is then in the ergodicity-breaking spin glass phase with exponentially
many equilibrium macroscopic states (complexity $\Sigma \neq 0$). The critical
inverse temperature $\beta_d$, at which the complexity $\Sigma$ starts to
deviate from zero, marks the onset of the spin glass phase.
This threshold quantity $\beta_d$ is referred to as 
the clustering or dynamical transition point in the literature
\cite{Mezard-Montanari-2009}.

\begin{figure}
  \includegraphics[width=0.45\textwidth]{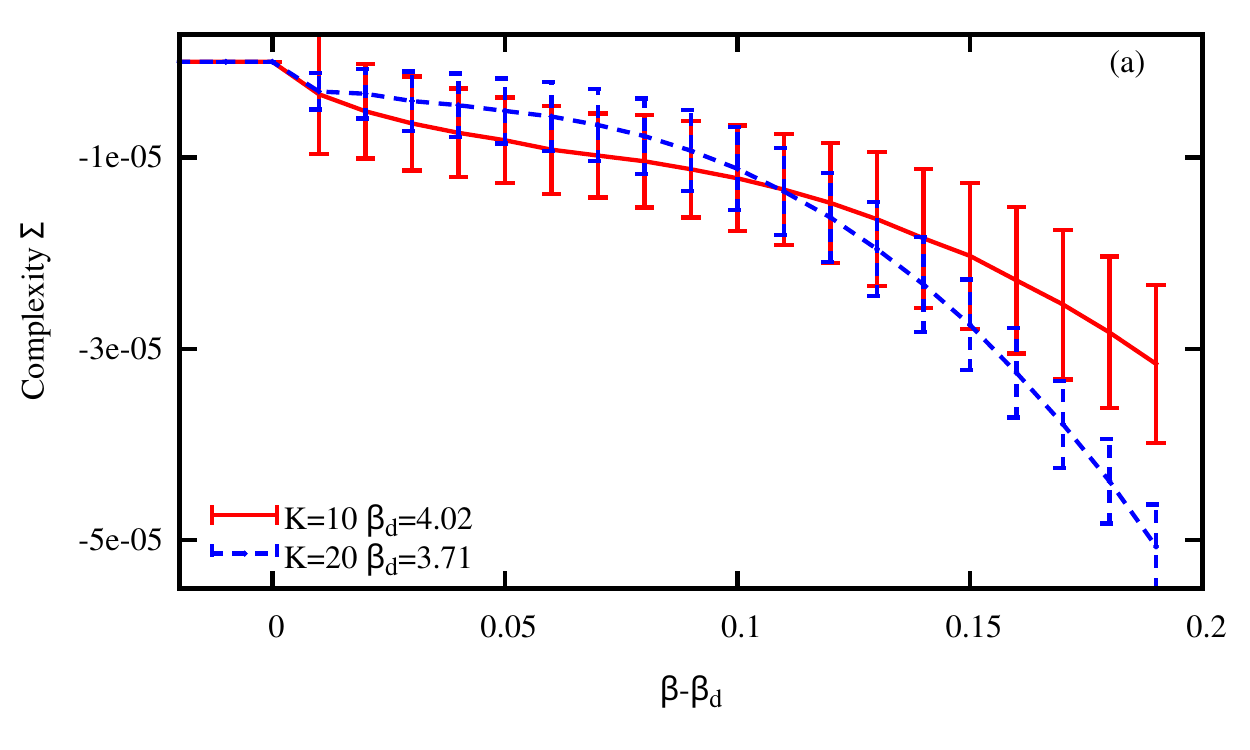}
  \includegraphics[width=0.45\textwidth]{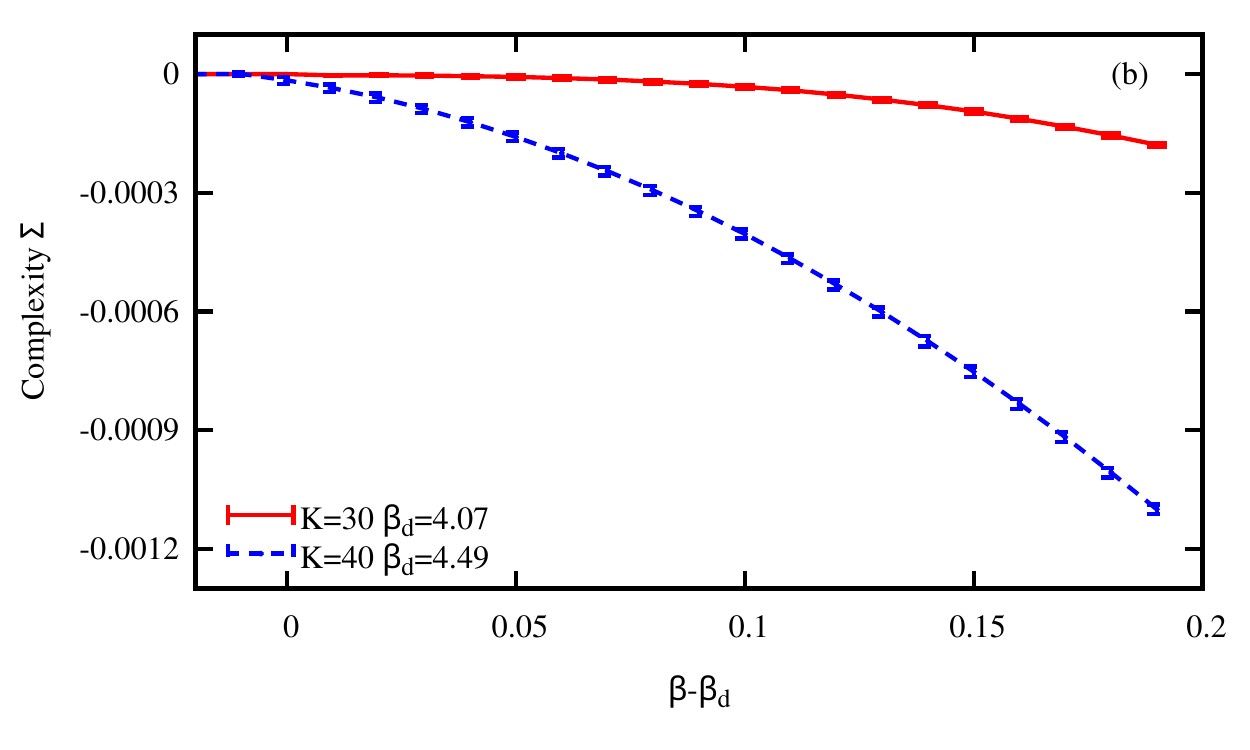}
  \includegraphics[width=0.45\textwidth]{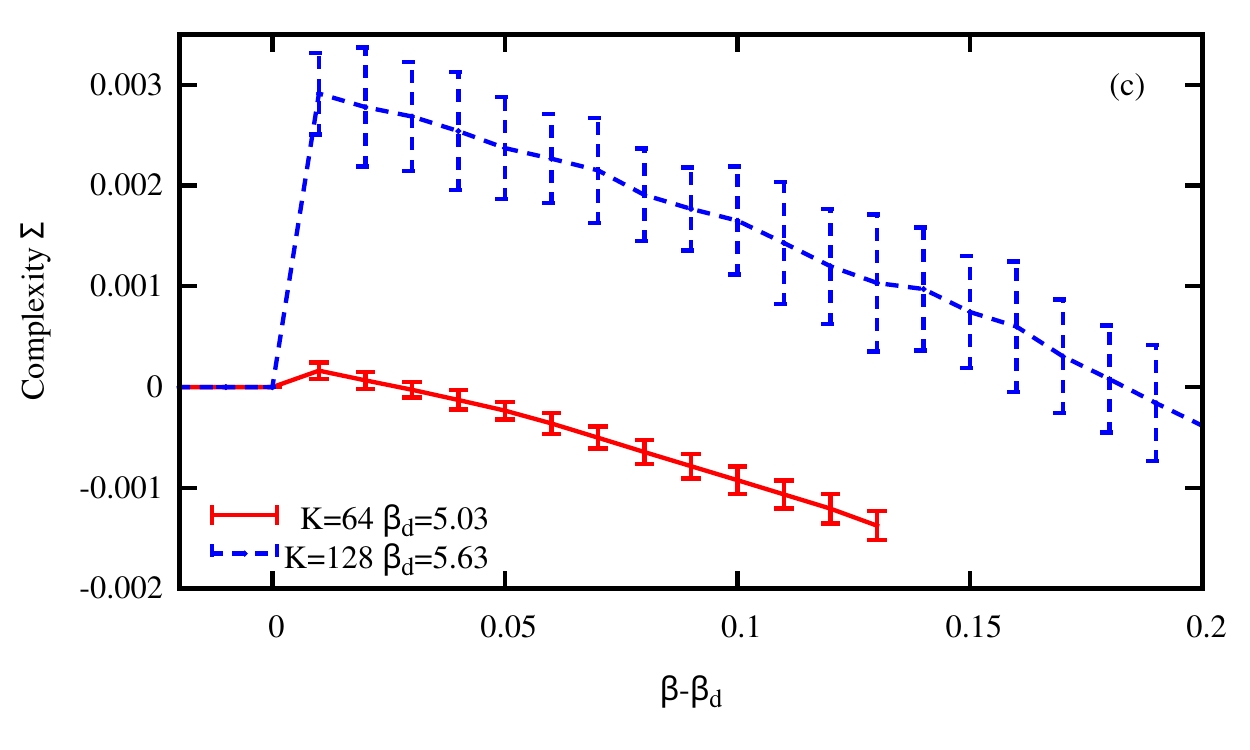}
  \caption{
    \label{fig:Complexity}
    The complexity $\Sigma$ of the regular random graph ensemble around the
    dynamical transition inverse temperature $\beta_d$. The vertex degrees $K$
    of the graph ensemble are:  (a) $10$ and $20$,
    (b) $30$ and $40$, and (c) $K=64$ and $128$.
  }
\end{figure}

\subsection{Critical inverse temperatures $\beta_d$ and $\beta_c$}

We can determine the ensemble-averaged complexity value $\Sigma$ as a function
of the inverse temperature $\beta$ (for $y=\beta$) 
by iterating Eq.~(\ref{eq:QcondBP}) through population dynamics
\cite{MEZ2001,PhysRevE.66.056126,Florent2007,Montanari2008,LenkaThe}. The major
numerical details are given in Appendix~\ref{sec:Popdynyb1}, and here we
describe the main results obtained by this method on the ensembles of RR
and ER graphs.

The dynamical transition inverse temperature $\beta_d$ as a function of the
vertex degree $K$ (for RR graphs) or mean vertex degree $c$ (for ER graphs)
is shown in Fig.~\ref{fig:PTpoint}. We find that $\beta_d$ is not a monotonic
function. For RR graphs, $\beta_d$ first decreases with $K$ and reaches
the minimum value of $\beta_d = 3.64 $ at $K = 16$,
then $\beta_d$ increases slowly with $K$. For $K<40$ the value of $\beta_d$
and the local stability inverse temperature $\beta_l$ are indistinguishable, 
but when $K>40$  the value of $\beta_l$ is noticeably higher than the value of
$\beta_d$. Similar situation occurs for ER graphs, for which $\beta_d$ reaches
the lowest value at $c\approx 57$, and $\beta_l$ becomes larger than
$\beta_d$ at $c>120$.

\begin{figure}
  \includegraphics[width=0.5\textwidth]{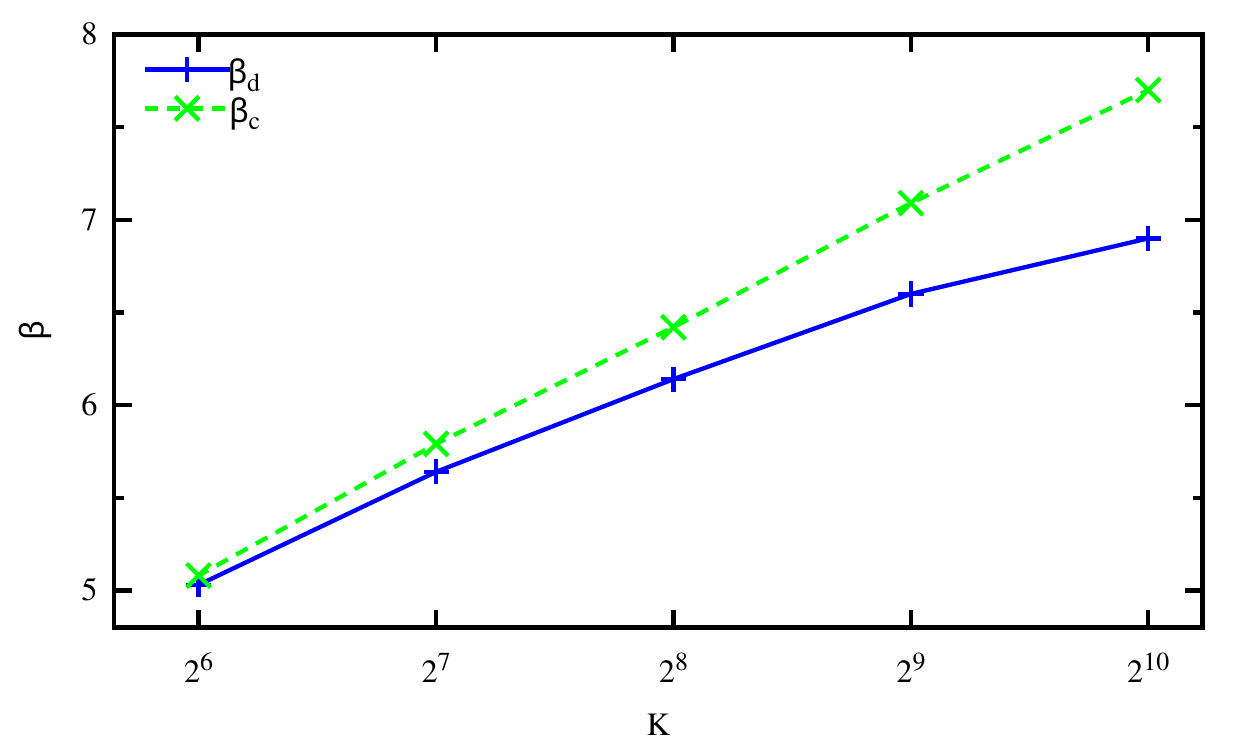}
  \caption{
    \label{fig:Td_Tk}
    The dynamical transition inverse temperature $\beta_d$ (plus symbols) 
    and the condense transition inverse temperature $\beta_c$ (cross symbols)
    for the regular random graph ensemble. The gap between $\beta_c$ and 
    $\beta_d$ increases with the vertex degree $K$ for $K\geq 64$, 
    while $\beta_d=\beta_c$ for $K<64$.
  }
\end{figure}

After the dynamical transition ($\beta > \beta_d$) the change of the
complexity in the vicinity of $\beta_d$ is shown in Fig.~\ref{fig:Complexity}
for several representative values of $K$ (RR graphs).
When $K < 64$ we find that $\Sigma$ is negative at $\beta > \beta_d$,
indicating that the equilibrium configuration space is dominated by only a
few macroscopic states. When $K \geq 64$, however, we
find that $\Sigma$ jumps from zero to a positive value at $\beta_d$ and
then gradually decreases with $\beta$, and it again becomes negative as $\beta$
exceeds a larger threshold value $\beta_c$. 
At each inverse temperature of $\beta \in (\beta_d, \beta_c)$ the equilibrium
configuration space is then contributed equally by an exponential number
($\approx e^{N \Sigma}$) of macroscopic states. The number of such macroscopic
states reduces to be $O(1)$ as the inverse temperature $\beta$ increases to
a larger threshold value $\beta_c$ (the condensation or the static phase
transition point), at which the the complexity $\Sigma$ computed at $y=\beta$
changes from being positive to being negative.
From Fig.~\ref{fig:Complexity}(c) and Fig.~\ref{fig:Td_Tk} we see that the gap
between $\beta_c$ and $\beta_d$ enlarges with the vertex degree $K$ (for
$K\geq 64$).

\begin{figure}
  \includegraphics[width=0.45\textwidth]{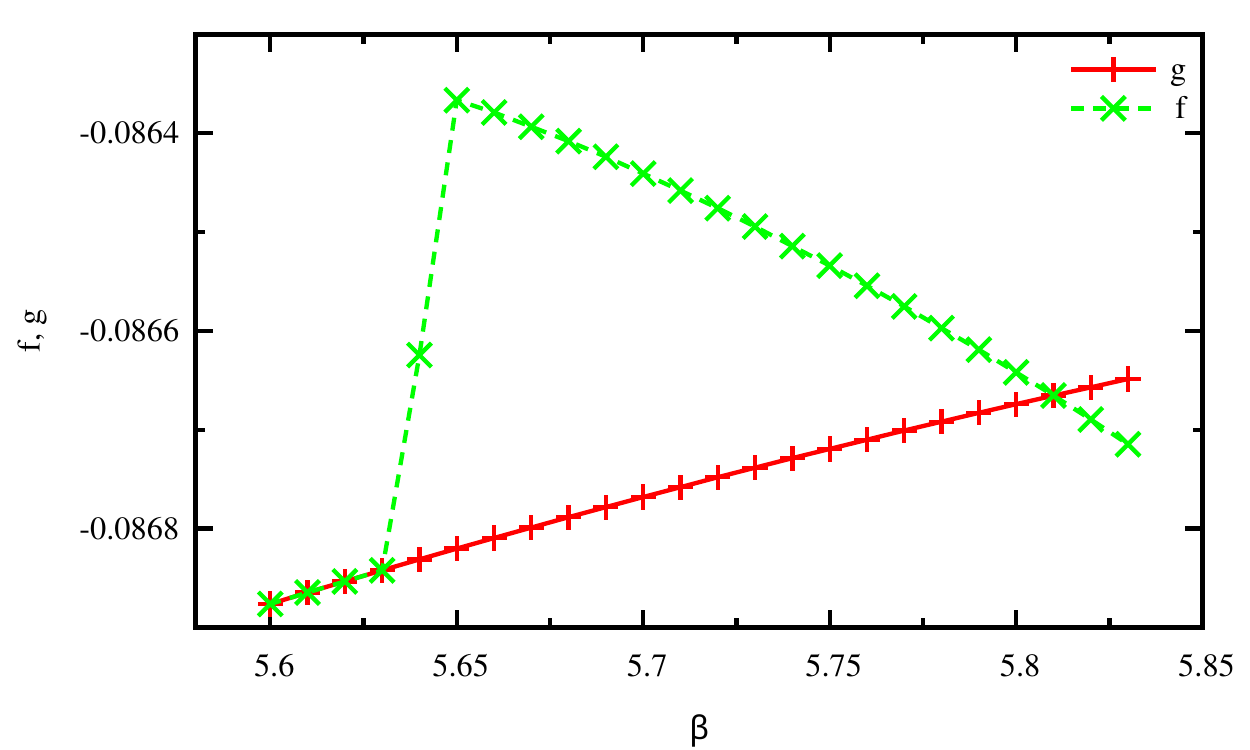}
  \caption{
    \label{fig:FGvsBeta}
    The mean free energy density ($\langle f \rangle$, cross symbols)
    and the grand free energy density ($g$, plus symbols)
    of the regular random graph ensemble at vertex degree $K=128$. 
    For this system $\beta_d \approx 5.63$ and $\beta_c \approx 5.81$.
  }
\end{figure}

For $\beta< \beta_d$ the system has only a unique equilibrium
macroscopic state, therefore the complexity $\Sigma=0$ and the mean free
energy density $\langle f \rangle$ is identical to the grand free energy
density $g$ (see Fig.~\ref{fig:FGvsBeta} for the case of $K=128$). At
$\beta=\beta_d$ the complexity $\Sigma$ jumps to a positive value and
the equilibrium configuration space breaks into $O(e^{N \Sigma})$ 
clusters (macroscopic states) with mean free energy density $\langle f \rangle$
larger than the grand free energy density $g$.
Notice that the grand free energy density $g$ changes smoothly at $\beta_d$
while $\langle f \rangle$ has a discontinuity (Fig.~\ref{fig:FGvsBeta}).
In the interval of  $\beta_d < \beta < \beta_c$, the mean free energy density 
$\langle f \rangle$ decreases with $\beta$ and the grand free energy density
$g$ increases with $\beta$, and $\langle f \rangle$ is equal to $g$ again
as $\beta$ reaches $\beta_c$.

We have also investigated the ER graph ensemble by the same method. 
We found that the complexity $\Sigma$ does not jump to a positive value at
the dynamical transition point $\beta_d$. Instead $\Sigma$ becomes negative
as $\beta$ increases from $\beta_d$ for all the considered mean vertex degree
$c$ values up to $c=512$. At the moment we could not
exclude the possibility that for $c$ sufficiently large the inverse
temperature $\beta_c$ will be distinct from the inverse temperature
$\beta_d$.

\section{The minimum feedback vertex set size}
\label{sec:T01RSB}

At each inverse temperature $\beta > \beta_c$, only a very few 
macroscopic states (those with the lowest free energy density)
are important to understand the equilibrium property of
the system. In this section let us consider the limiting case
of $\beta\rightarrow \infty$ which corresponds to the
minimum FVS problem. At this limit the 1RSB mean field theory can be
simplified to a considerable exent
\cite{Alfredo2004,Alfredo2002, PhysRevE.66.056126, PhysRevE.77.066102, ZhouJie2007}. The corresponding message-passing algorithm at finite value of
$y$  is referred to as SP(y), i.e., survey propagation at finite $y$.

Before deriving the SP(y) mean field equations, we first need to
obtain the zero-temperature limit of the BP equation
(\ref{eq:BP}). This limit is also known as the max-product or
min-sum algorithm \cite{Alfredo2004,MEZ2001,FlorentEPL2008}.
It is convenient for us to rewrite the cavity messages as the power of 
cavity fields:
\begin{subequations}
  \begin{align}
    q_{i\rightarrow j}^0 & = e^{-\beta \gamma_{i\rightarrow j}} \; , \\
    q_{i\rightarrow j}^i & = e^{-\beta \rho_{i\rightarrow j}} \; , \\
    q_{i\rightarrow j}^X & = e^{-\beta \eta_{i\rightarrow j}} \; .
  \end{align}
\end{subequations}
At the limit of $\beta \rightarrow \infty$, we obtain from Eq.~(\ref{eq:BP})
the following iterative equations for $\gamma_{i\rightarrow j}$,
$\rho_{i\rightarrow j}$, and $\eta_{i\rightarrow j}$:
\begin{subequations}
  \label{eq:min-sum}
  \begin{align}
   & \gamma_{i\rightarrow j}  = 1.0 -\min\biggl( 1.0, \, 
    \sum_{k\in\partial i\setminus j} 
    \min(\gamma_{k\rightarrow i}, \rho_{k\rightarrow i}), \nonumber\\
    & \quad \sum_{k\in\partial i\setminus j} \min(\gamma_{k\rightarrow i},
    \rho_{k\rightarrow i})-\min\Bigl(\bigl\{\min(\rho_{l\rightarrow i},
    \eta_{l\rightarrow i}) \nonumber\\
    & \quad -\min(\gamma_{l\rightarrow i},
    \rho_{l\rightarrow i})\bigr\}_{l\in\partial i\setminus j,k}\Bigr) \biggr) \; , \\
    & \rho_{i\rightarrow j}  = 
    \gamma_{i\rightarrow j}-1.0+
    \sum_{k\in\partial i\setminus j}
    \min(\gamma_{k\rightarrow i},\rho_{k\rightarrow i}) \; ,\\
    & \eta_{i\rightarrow j}  = 
    \gamma_{i\rightarrow j}-1.0+\sum_{k\in\partial i\setminus j}
    \min(\gamma_{k\rightarrow i},\rho_{k\rightarrow i})\nonumber\\
    & +\min\Bigl(\bigl\{\min(\rho_{l\rightarrow i},\eta_{l\rightarrow i})
    -\min(\gamma_{l\rightarrow i},\rho_{l\rightarrow i})
    \bigr\}_{l\in\partial i\setminus j,k} \biggr) \; .
    \nonumber\\ 
    \label{eq:eta}
  \end{align}
\end{subequations}
Notice that $\gamma_{i\rightarrow j} \in [0,1]$ while the values of
$\eta_{i\rightarrow j}$ and $\rho_{i\rightarrow j}$ can be greater than $1.0$, and
furthermore if $\gamma_{i\rightarrow j} >0$ then either $\rho_{i\rightarrow j}=0$
or $\eta_{i\rightarrow j}=0$.
To simplify the notation we denote by 
$\chi_{i\rightarrow j} \equiv 
\{\gamma_{i\rightarrow j},\rho_{i\rightarrow j}, \eta_{i\rightarrow j}\}$
the three cavity field messages from vertex $i$ to vertex $j$. The
min-sum BP equation (\ref{eq:min-sum}) is then denoted as
$\chi_{i\rightarrow j}= b_{e}\bigl(\{\chi_{k\rightarrow i}: k\in
\partial i\setminus j\}\bigr)$.

At the $\beta\rightarrow \infty$ the free energy contributions $f_i$ and
$f_{i j}$ of a vertex $i$ and an edge have the corresponding limiting
value $\hat{f}_i$ and $\hat{f}_{i j}$:
\begin{subequations}
  \begin{align}
    & \hat{f}_i = -\max\biggl(0, \, 1.0 -
    \sum_{k\in\partial i} 
    \min(\gamma_{k\rightarrow i},\rho_{k\rightarrow i}), \nonumber \\
    & \quad
    1.0-\sum_{k\in\partial i}\min(\gamma_{k\rightarrow i},\rho_{k\rightarrow i})
    -\min\Bigl(\bigl\{\min(\rho_{l\rightarrow i},\eta_{l\rightarrow i})
    \nonumber \\
    & \quad \quad 
    -\min(\gamma_{l\rightarrow i},\rho_{l\rightarrow i})
    \bigr\}_{l\in\partial i\setminus k}\Bigr) \biggr) \; , \\
    & \hat{f}_{i j} = \min\biggl( (\gamma_{i\rightarrow j}+\gamma_{j\rightarrow i}),
    \nonumber\\
    & \quad \min(\rho_{i\rightarrow j},\eta_{i\rightarrow j})+
    \min(\rho_{j\rightarrow i},\gamma_{j\rightarrow i}),\nonumber\\
    & \quad \min(\rho_{j\rightarrow i},\eta_{j\rightarrow i})+
    \min(\rho_{i\rightarrow j},\gamma_{i\rightarrow j}) \biggr) \; .
  \end{align}
\end{subequations}

At $\beta\rightarrow \infty$ the probability functional 
$Q_{i\rightarrow j}[q_{i\rightarrow j}]$ of Eq.~(\ref{eq:1RSB})
correponds to the probability function
$\Theta_{i\rightarrow j}[\chi_{i\rightarrow j}]$, and the self-consistent
equation for this function is
\begin{equation}
  \Theta_{i\rightarrow j}[\chi]= 
  \frac{\int \prod\limits_{k\in \partial i\setminus j} 
    \mathcal{D}\chi_{k \rightarrow i}\Theta_{k \rightarrow i}[\chi] 
    e^{-y \hat{f}_{i\rightarrow j}}\delta\bigl[\chi-b_{e}(\{\chi\})\bigr]}{\int 
    \prod\limits_{k\in \partial i\setminus j} \mathcal{D}\chi_{k \rightarrow i}
    \Theta_{k \rightarrow i}[\chi]e^{-y \hat{f}_{i\rightarrow j}}}\; ,
\end{equation}
where $\hat{f}_{i\rightarrow j}=-\gamma_{i\rightarrow j}$.
The $\beta\rightarrow \infty$ grand free energy
$\mathbb{G}=\sum_i\mathbb{G}_i-\sum_{(i,j)}\mathbb{G}_{ij}$ and the mean free 
energy $\mathbb{F}=\sum_i\mathbb{F}_i-\sum_{(i,j)}\mathbb{F}_{ij}$ are obtained
by computing
\begin{eqnarray}
  \mathbb{G}_i&=&-\frac{1}{y}\log\bigg[\int \mathcal{D} 
    \chi_{j\rightarrow i} \prod_{j\in\partial i}
    \Theta_{j\rightarrow i}[\chi]e^{-y\hat{f}_i}\bigg] \; ,\\
  \mathbb{G}_{ij}&=&-\frac{1}{y}\log\bigg[\int \mathcal{D} 
    \chi_{j\rightarrow i} \mathcal{D} \chi_{i\rightarrow j}
    \Theta_{j\rightarrow i}[\chi]\Theta_{i\rightarrow j}[\chi]
    e^{-y\hat{f}_{ij}}\bigg] \; ,\nonumber\\
  \\
  \mathbb{F}_i&=&-\frac{\int \mathcal{D} \chi_{j\rightarrow i} 
    \prod_{j\in\partial i}\Theta_{j\rightarrow i}[\chi]\hat{f}_i
    e^{-y\hat{f}_i}}{\int \mathcal{D} \chi_{j\rightarrow i} 
    \prod_{j\in\partial i}\Theta_{j\rightarrow i}[\chi]e^{-y\hat{f}_i}} \; ,\\
  \mathbb{F}_{ij}&=&-\frac{\int \mathcal{D} \chi_{j\rightarrow i}
    \mathcal{D} \chi_{i\rightarrow j}\Theta_{j\rightarrow i}[\chi]
    \Theta_{i\rightarrow j}[\chi]\hat{f}_{ij}e^{-y\hat{f}_{ij}}}{\int \mathcal{D} 
    \chi_{j\rightarrow i} \mathcal{D} \chi_{i\rightarrow j}
    \Theta_{j\rightarrow i}[\chi]\Theta_{i\rightarrow j}[\chi]e^{-y\hat{f}_{ij}}} \; .
\end{eqnarray}
The $\beta\rightarrow \infty$ complexity, $\Sigma^e$ is then computed as
$\Sigma^e=y\bigl(\mathbb{F}-\mathbb{G} \bigr)/N$.

\begin{figure}
  \includegraphics[width=0.45\textwidth]{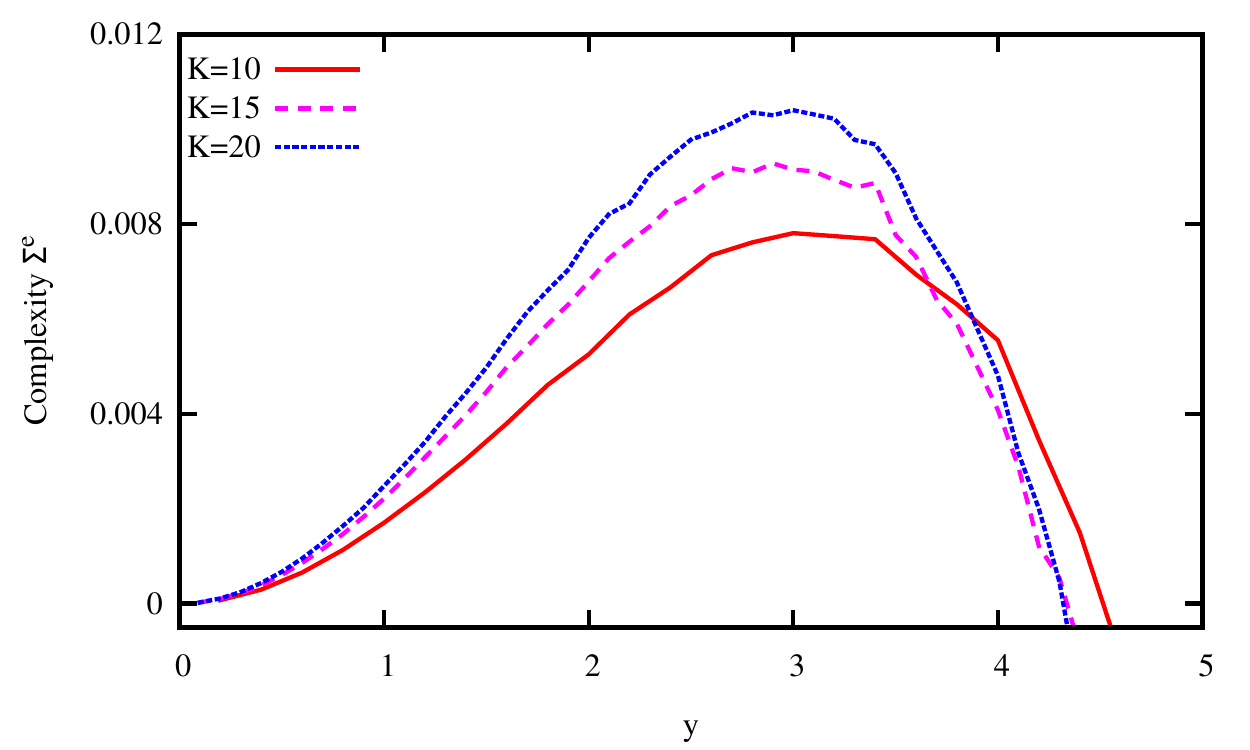}
  \caption{
    \label{fig:SPy}
    The $\beta\rightarrow \infty$ complexity $\Sigma^e$ of the regular
    random graph ensemble as a function of the reweighting parameter $y$.
    The vertex degrees are $K=10$, $15$, and $30$ for the three curves.
  }
\end{figure}

\begin{figure}
  \includegraphics[width=0.45\textwidth]{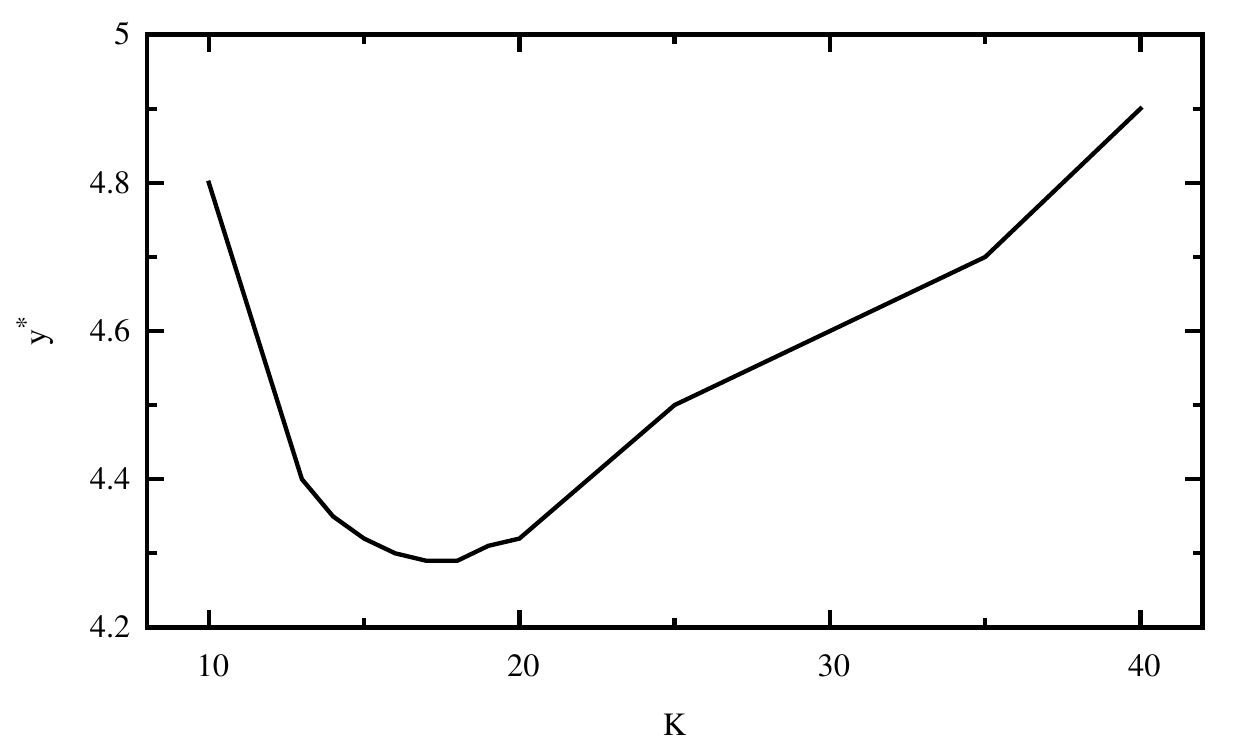}
  \caption{
    \label{fig:SPyS0y}
    The relationship between the value of $y^*$, at which the 
    complexity $\Sigma^e$ changes from being
    positive to being negative, and the vertex degree $K$ of the regular
    random graph ensemble.
  }
\end{figure}

The RR ensemble-averaged complexity $\Sigma^e$ as a function of $y$ is shown
in Fig.~\ref{fig:SPy} for several different values of vertex degree $K$.
The computation details and the pseudocode are explained in
Appendix~\ref{sec:appSPy}. We see that $\Sigma^e$ first increases from zero
with $y$ and it reaches a maximum at $y \sim 3$ and then decreases to below
zero. The solution of $\Sigma^e(y)=0$ is denoted by $y^*$, which corresponds
to the minimum energy density $e_0$ (i.e., the free energy density
$\mathbb{F}/N$ computed at $y=y^*$). Figure \ref{fig:SPyS0y} demonstrates that
$y^*$ is not a monotonic function of the vertex degree $K$, and the minimum
value of $y^*$ is reached at $K=17$.

\begin{figure}
  \includegraphics[width=0.45\textwidth]{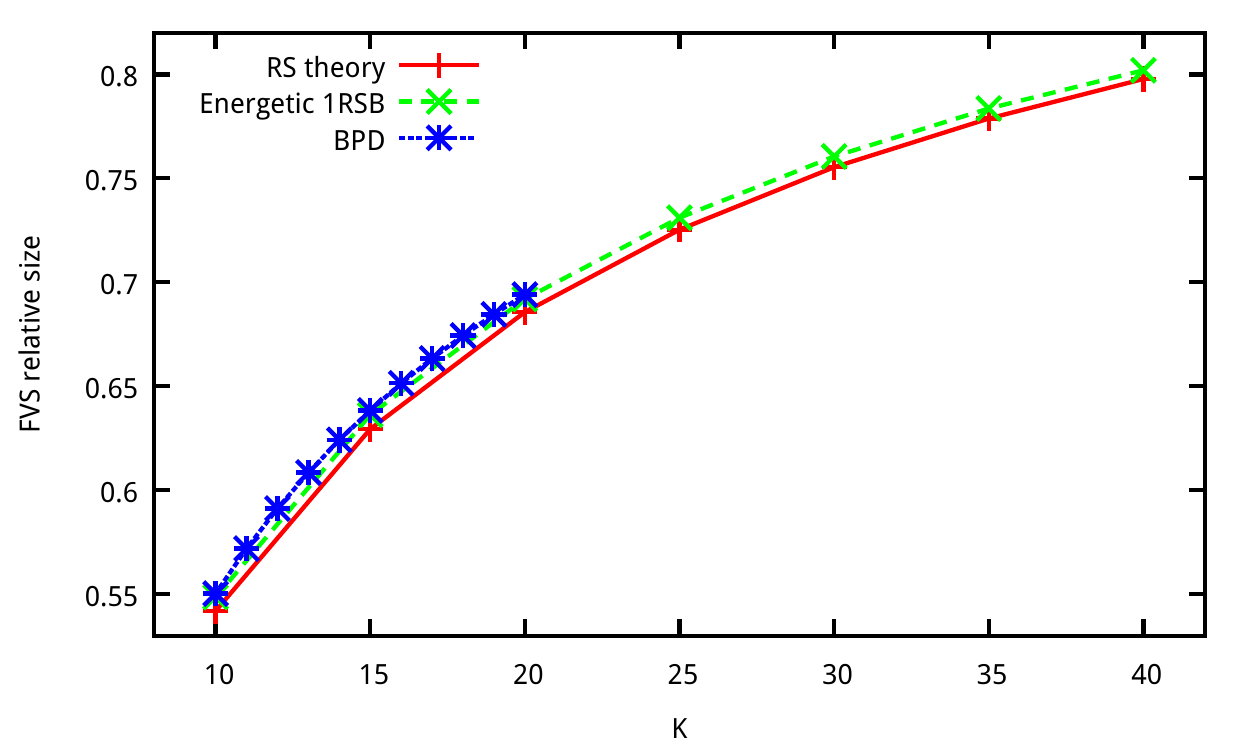}
  \caption{
    \label{fig:FVS_size}
    The relative sizes of minimum feedback vertex sets as predicted by the
    $\beta\rightarrow \infty$ 1RSB mean field theory  (cross symbols)
    and the RS mean field theory (plus symbols), and the relative sizes of
    feedback vertex sets obtained by the BPD algorithm of \cite{HJZEPJB2013}
    (star symbols). $K$ is the vertex degree of regular random graphs.
  }
\end{figure}

The RR ensemble-averaged minimum energy density $e_0$ (which is the relative
size of a minimum FVS) is compared with the prediction of the RS mean field
theory in Fig.~\ref{fig:FVS_size}. We notice that the 1RSB prediction is
higher than the RS prediction, and it is in much better agreement with the
results of the belief propagation-guided decimation (BPD) algorithm, which
constructs close-to-minimum feedback vertex sets for single graph instances
\cite{HJZEPJB2013}. The 1RSB energetic results therefore confirm  that (1) 
the minimal FVS cardinality of a RR graph is indeed higher than the value 
predicted by the RS mean field theory \cite{HJZEPJB2013} and that (2) the
BPD algorithm is very efficient for the RR graph ensemble (its efficiency
for the ER graph ensemble has already been confirmed in \cite{HJZEPJB2013}). 
 
Based on the iterative equation (\ref{eq:min-sum}) we can easily implement a
survey propagation-guided decimation (SPD) algorithm as a solver for the
minimal FVS problem. The implementation details are largely the same as those
of the BPD algorithm \cite{HJZEPJB2013}. Given that the BPD algorithm is
already excellent for RR and ER random graphs, the improvement of SPD over
BPD is expected to be insignificant for these two graph ensembles. But for
some real-world network instances with complicated structural correlations
SPD might achieve better performance than BPD.

\section{Conclusion}

In this paper we studied the stability of the Replica-symmetric
mean field theory of the undirected FVS problem, and investigated the
low-temperature energy landscape property of this problem by the 1RSB
mean field theory. We determined the dynamical (clustering) phase transition
inverse temperature $\beta_d$ and the static (condensation) 
phase transition inverse
temperature $\beta_c$ for both RR and ER random graph ensembles, and we
also computed the minimum FVS size of the RR graph ensemble by the 
$\beta\rightarrow \infty$ 1RSB mean field theory.

One of our major theoretical results is that, for the RR graph ensemble
with vertex degree $K\geq 64$, the undirected FVS problem has two distinct
phase transitions, one clustering transition at inverse temperature
$\beta=\beta_d$ followed by another condensation transition at a higher
$\beta=\beta_c$. The existence of two separate spin glass transitions is a
common feature for  many-body interaction models (like the random 
$K$-satisfiability
problem with $K \geq 4$ \cite{Montanari2008} and the $p$-spin glass model
with $p\geq 3$ \cite{PhysRevLett.87.127209,franz2001}) and many-states systems 
(like the $Q$-coloring problem with $Q\geq 4$
\cite{FlorentEPL2008}). Such a feature has also been predicted to occur in
the random vertex cover problem
\cite{PhysRevE.80.021122,CojaOghlan-Efthymiou-2014}. 

For the ER graph ensemble up to mean vertex degree $c=512$ our 1RSB mean
field theory predicts that $\beta_c = \beta_d$. Maybe $c$ needs to be 
very large for $\beta_c$ to be distinct from $\beta_d$ for this graph ensemble.
One way of checking this possibility is to study the 1RSB mean field theory
at the large $c$ limit \cite{ZhangPan2013,PhysRevE.56.1357}, but we have not
yet carried out such an effort in this paper.

\section*{Acknowledgement}

The authors thank Chuang Wang, Pan Zhang and Jin-Hua Zhao for helpful
discussions. YZ thanks Professor Fangfu Ye for encouragement and support.
The computations in this paper are carried out in the
HPC computing cluster of ITP-CAS. HJZ was supported by the National Basic
Research Program of China (grant number 2013CB932804) and by the National
Natural Science Foundation of China (grand numbers 11121403 and 11225526).
SMQ was partially supported by the
Scientific Research Foundation of CAUC (grant number 2015QD093).

\widetext

\appendix

\section{Computing thermodynamical quantities at $y / \beta=1$}
\label{sec:appyb1}

According to the 1RSB mean field theory, the grand free energy density $g$ for
a given graph $\mathcal{G}$ is computed through
\begin{equation}
  \label{eq:g1RSB}
  g  = \frac{1}{N} (\sum_{i=1}^{N} g_i - \sum_{(i,j)\in \mathcal{G}} g_{ij}) \; .
\end{equation}
At $y=\beta$, the grand free energy contributions of a vertex $i$ and an edge
$(i, j)$ can be evaluated by the following simplified expressions
\begin{subequations}
  \begin{align}
    g_i & = -\frac{1}{\beta}\ln\Bigl[ e^{-\beta}+\prod\limits_{j \in \partial i}
      \bigl[\overline{q}^0_{j\rightarrow i}+\overline{q}^j_{j \rightarrow i} \bigr]
      +  \sum\limits_{j\in \partial i} (1- \overline{q}_{j\rightarrow i}^0)
      \prod\limits_{k\in \partial i\backslash j} \bigl[
        \overline{q}^0_{k\rightarrow i}+\overline{q}^k_{k \rightarrow i}\bigr]
      \Bigr] \; , \\
    g_{ij} & =  -\frac{1}{\beta}\ln \Bigl[\overline{q}^0_{i\rightarrow j}
      \overline{q}^0_{j\rightarrow i} + (1-\overline{q}^0_{i\rightarrow j})
      (\overline{q}^0_{j\rightarrow i}+\overline{q}^j_{j\rightarrow i})
      + (1-\overline{q}^0_{j\rightarrow i})(\overline{q}^0_{i\rightarrow j}+
      \overline{q}^i_{i\rightarrow j}) \Bigr] \; .
  \end{align}
\end{subequations}

Similar to Eq.~(\ref{eq:g1RSB}), the mean free energy density
$\langle f \rangle$ of a macroscopic state is computed through
\begin{equation}
  \langle f \rangle =  \frac{1}{N} (\sum_{i=1}^{N} \langle f_i \rangle -
 \sum_{(i,j)\in \mathcal{G}} \langle f_{ij} \rangle \;) .
\end{equation}
Then the complexity $\Sigma$ of the system at fixed values of $\beta$ and $y$
is evaluated through
\begin{equation}
  \Sigma = y \bigl( \langle f \rangle - g \bigr) \; .
\end{equation}

The mean free energy contribution $\langle f_i \rangle$ of a
vertex $i$ can be computed through
\begin{equation}
  \langle f_i \rangle  = \overline{q}_i^0 \langle f_i^0 \rangle +
  \overline{q}_i^i \langle f_i^i \rangle + \overline{q}_i^X \langle
  f_i^X \rangle \; .
\end{equation}
In the above expression, $\overline{q}_i^0$, $\overline{q}_i^i$, and
$\overline{q}_i^X$ are, respectively, the mean value of $q_{i}^0$,
$q_i^i$ and $q_i^X$ over all the macroscopic states:
\begin{subequations}
  \begin{align}
    \overline{q}^0_i & =  \frac{e^{-\beta}}{
      e^{-\beta} + \prod\limits_{j\in \partial i}
      \bigl[\overline{q}_{j\rightarrow i}^0 +\overline{q}_{j\rightarrow i}^j \bigr]
      + \sum\limits_{k\in \partial i} (1-\overline{q}_{k\rightarrow i}^0)
      \prod\limits_{j\in \partial i\backslash k} \bigl[
        \overline{q}_{j\rightarrow i}^0 + \overline{q}_{j\rightarrow i}^j \bigr]
    } \; ,
    \\
    \overline{q}^i_i & =  \frac{\prod_{j \in \partial i}
      \bigl[\overline{q}^0_{j\rightarrow i}+\overline{q}^j_{j \rightarrow i}\bigr]}
             { e^{-\beta} + \prod\limits_{j\in \partial i}
      \bigl[\overline{q}_{j\rightarrow i}^0 +\overline{q}_{j\rightarrow i}^j \bigr]
      + \sum\limits_{k\in \partial i} (1-\overline{q}_{k\rightarrow i}^0)
      \prod\limits_{j\in \partial i\backslash k} \bigl[
        \overline{q}_{j\rightarrow i}^0 + \overline{q}_{j\rightarrow i}^j \bigr]
    } \; ,
    \\
    \overline{q}^X_i & =  \frac{
      \sum\limits_{k\in \partial i} (1-\overline{q}_{k\rightarrow i}^0)
      \prod\limits_{j\in \partial i\backslash k} \bigl[
        \overline{q}_{j\rightarrow i}^0 + \overline{q}_{j\rightarrow i}^j \bigr]
    }{
      e^{-\beta} + \prod\limits_{j\in \partial i}
      \bigl[\overline{q}_{j\rightarrow i}^0 +\overline{q}_{j\rightarrow i}^j \bigr]
      + \sum\limits_{k\in \partial i} (1-\overline{q}_{k\rightarrow i}^0)
      \prod\limits_{j\in \partial i\backslash k} \bigl[
        \overline{q}_{j\rightarrow i}^0 + \overline{q}_{j\rightarrow i}^j \bigr]
    } \; ,
  \end{align}
\end{subequations}
while the explicit expressions for $\langle f_i^0 \rangle$, $\langle f_i^i \rangle$ and $\langle f_i^X \rangle$ are
\begin{subequations}
  \begin{align}
    \langle f_i^0 \rangle & =
    \prod\limits_{j \in \partial i} \int \mathcal{D} {\bf q}_{j\rightarrow i}\,
    \biggl[\overline{q}_{j\rightarrow i}^0 \, Q_{j\rightarrow i}^0 \bigl[
        {\bf q}_{j\rightarrow i} \bigr] +  \overline{q}_{j\rightarrow i}^j
      \, Q_{j\rightarrow i}^j \bigl[
        {\bf q}_{j\rightarrow i} \bigr] + \overline{q}_{j\rightarrow i}^X \,
      Q_{j\rightarrow i}^X \bigl[ {\bf q}_{k\rightarrow i} \bigr] \biggr] \,
    f_i\bigl(\{q_{m\rightarrow i}\}\bigr) \; , \\
    \langle f_{i}^i \rangle & =  \prod\limits_{j\in
      \partial i} \int \mathcal{D}{\bf q}_{j\rightarrow i} \biggl[
      \frac{\overline{q}_{j\rightarrow i}^0} {\overline{q}_{j \rightarrow i}^0+
        \overline{q}_{j \rightarrow i}^j} Q_{j\rightarrow i}^0 \bigl[
        {\bf q}_{j\rightarrow i}\bigr] + \frac{\overline{q}_{j\rightarrow i}^j}{
        \overline{q}_{j \rightarrow i}^0+ \overline{q}_{j \rightarrow i}^j}
      Q_{j\rightarrow i}^j\bigl[ {\bf q}_{k\rightarrow i}\bigr] \biggr] \,
  f_i\bigl(\{q_{m\rightarrow i}\}\bigr) \; , \\
  \langle f_i^X \rangle & = \sum_{j\in\partial i}\omega_j \int\mathcal{D}
        {\bf q}_{j\rightarrow i}
    \biggl[ \frac{\overline{q}_{j\rightarrow i}^j}{
        \overline{q}_{j \rightarrow i}^j + \overline{q}_{j \rightarrow i}^X}
      Q_{j\rightarrow i}^j \bigl[{\bf q}_{j\rightarrow i}\bigr] +
      \frac{\overline{q}_{j\rightarrow i}^X}{
        \overline{q}_{j \rightarrow i}^j + \overline{q}_{j \rightarrow i}^X}
      Q_{j\rightarrow i}^X \bigl[{\bf q}_{j\rightarrow i}\bigr] \biggr]
    \times \nonumber \\
    &\quad \quad \prod\limits_{k\in \partial i\backslash j}\int \mathcal{D}
           {\bf q}_{k\rightarrow i} \biggl[ \frac{
               \overline{q}_{k\rightarrow i}^0}
             {\overline{q}_{k \rightarrow i}^0 +
               \overline{q}_{k \rightarrow i}^k}
             Q_{k\rightarrow i}^0 \bigl[{\bf q}_{k\rightarrow i}\bigr] +
             \frac{\overline{q}_{k\rightarrow i}^k}{
               \overline{q}_{k \rightarrow i}^0 +
               \overline{q}_{k \rightarrow i}^k}
             Q_{k\rightarrow i}^k \bigl[{\bf q}_{k\rightarrow i}\bigr] \biggr]
           f_i\bigl(\{q_{m\rightarrow i}\}\bigr) \; ,
  \end{align}
\end{subequations}
with
\begin{equation}
  \omega_j = \frac{(1-\overline{q}_{j\rightarrow i}^0)
    \prod\limits_{k\in \partial i\backslash j} \bigl[
      \overline{q}_{k\rightarrow i}^0 + \overline{q}_{k\rightarrow i}^k \bigr]
  }{\sum\limits_{l\in \partial i} (1-\overline{q}_{l\rightarrow i}^0)
    \prod\limits_{k\in \partial i\backslash l} \bigl[
      \overline{q}_{k\rightarrow i}^0 + \overline{q}_{k\rightarrow i}^k \bigr]
  }  \; .
\end{equation}
The mean free energy contribution $\langle f_{ij} \rangle$ of
an edge $(i, j)$ can be computed through
\begin{equation}
  \langle f_{ij} \rangle = \omega_{ij}^0 \langle f_{ij}^0 \rangle
  + \omega_{ij}^i \langle f_{ij}^i \rangle + \omega_{ij}^j \langle
  f_{ij}^j \rangle \; ,
\end{equation}
where
\begin{subequations}
  \begin{align}
    \omega_{ij}^0 & = \frac{\overline{q}_{i\rightarrow j}^0
      \overline{q}_{j\rightarrow i}^0}{\overline{q}_{i\rightarrow j}^0
      \overline{q}_{j\rightarrow i}^0 +(1-\overline{q}_{i\rightarrow j}^0)
      (\overline{q}_{j\rightarrow i}^0 + \overline{q}_{j\rightarrow i}^j) +
      (1-\overline{q}_{j\rightarrow i}^0)
      (\overline{q}_{i\rightarrow j}^0 + \overline{q}_{i\rightarrow j}^i)
    }\; , \\
    \omega_{ij}^i & =
    \frac{(1- \overline{q}_{i\rightarrow j}^0) (\overline{q}_{j\rightarrow i}^0 +
      \overline{q}_{j\rightarrow i}^j)}{\overline{q}_{i\rightarrow j}^0
      \overline{q}_{j\rightarrow i}^0 +(1-\overline{q}_{i\rightarrow j}^0)
      (\overline{q}_{j\rightarrow i}^0 + \overline{q}_{j\rightarrow i}^j) +
      (1-\overline{q}_{j\rightarrow i}^0)
      (\overline{q}_{i\rightarrow j}^0 + \overline{q}_{i\rightarrow j}^i)
    }\; , \\
    \omega_{ij}^j & =
    \frac{(1- \overline{q}_{j\rightarrow i}^0) (\overline{q}_{i \rightarrow j}^0 +
      \overline{q}_{i\rightarrow j}^i)}{\overline{q}_{i\rightarrow j}^0
      \overline{q}_{j\rightarrow i}^0 +(1-\overline{q}_{i\rightarrow j}^0)
      (\overline{q}_{j\rightarrow i}^0 + \overline{q}_{j\rightarrow i}^j) +
      (1-\overline{q}_{j\rightarrow i}^0)
      (\overline{q}_{i\rightarrow j}^0 + \overline{q}_{i\rightarrow j}^i)
    }\; ,
  \end{align}
\end{subequations}
and
\begin{subequations}
  \begin{align}
    \langle f_{ij}^0 \rangle & = \int \mathcal{D} {\bf q}_{j\rightarrow i}\,
    Q_{j\rightarrow i}^0 \bigl[{\bf q}_{j\rightarrow i} \bigr]
    \int \mathcal{D} {\bf q}_{i\rightarrow j}\, Q_{i\rightarrow j}^0 \bigl[
      {\bf q}_{i\rightarrow j} \bigr] \,
    f_{ij}\bigl( {\bf q}_{i\rightarrow j}, {\bf q}_{j\rightarrow i}\bigr) \; , \\
    \langle f_{ij}^i \rangle & = \int \mathcal{D} {\bf q}_{i\rightarrow j}\,
    \biggl[ \frac{\overline{q}_{i\rightarrow j}^i}{
        \overline{q}_{i \rightarrow j}^i + \overline{q}_{i \rightarrow j}^X}
      Q_{i\rightarrow j}^i \bigl[{\bf q}_{i\rightarrow j}\bigr] +
      \frac{\overline{q}_{i\rightarrow j}^X}{
        \overline{q}_{i\rightarrow j}^i + \overline{q}_{i\rightarrow j}^X}
      Q_{i\rightarrow j}^X \bigl[{\bf q}_{i\rightarrow j}\bigr] \biggr] \times
    \nonumber \\
    & \quad \int \mathcal{D} {\bf q}_{j\rightarrow i} \biggl[
      \frac{\overline{q}_{j\rightarrow i}^0}{
        \overline{q}_{j\rightarrow i}^0 + \overline{q}_{j\rightarrow i}^j}
      Q_{j\rightarrow j}^0 \bigl[{\bf q}_{j\rightarrow i}\bigr] +
      \frac{\overline{q}_{j\rightarrow i}^j}{
        \overline{q}_{j\rightarrow i}^0 + \overline{q}_{j\rightarrow i}^j}
      Q_{j\rightarrow i}^j \bigl[{\bf q}_{j\rightarrow i}\bigr] \biggr] \,
    f_{ij}\bigl( {\bf q}_{i\rightarrow j}, {\bf q}_{j\rightarrow i}\bigr) \; , \\
    \langle f_{ij}^j \rangle & = \int \mathcal{D} {\bf q}_{i\rightarrow j}\,
    \biggl[ \frac{\overline{q}_{i\rightarrow j}^0}{
        \overline{q}_{i \rightarrow j}^0 + \overline{q}_{i \rightarrow j}^i}
      Q_{i\rightarrow j}^0 \bigl[{\bf q}_{i\rightarrow j}\bigr] +
      \frac{\overline{q}_{i\rightarrow j}^i}{
        \overline{q}_{i\rightarrow j}^0 + \overline{q}_{i\rightarrow j}^i}
      Q_{i\rightarrow j}^i \bigl[{\bf q}_{i\rightarrow j}\bigr] \biggr] \times
    \nonumber \\
    & \quad \int \mathcal{D} {\bf q}_{j\rightarrow i} \biggl[
      \frac{\overline{q}_{j\rightarrow i}^j}{
        \overline{q}_{j\rightarrow i}^j + \overline{q}_{j\rightarrow i}^X}
      Q_{j\rightarrow j}^j \bigl[{\bf q}_{j\rightarrow i}\bigr] +
      \frac{\overline{q}_{j\rightarrow i}^X}{
        \overline{q}_{j\rightarrow i}^j + \overline{q}_{j\rightarrow i}^X}
      Q_{j\rightarrow i}^X \bigl[{\bf q}_{j\rightarrow i}\bigr] \biggr] \,
    f_{ij}\bigl( {\bf q}_{i\rightarrow j}, {\bf q}_{j\rightarrow i}\bigr) \; .
  \end{align}
\end{subequations}

\section{1RSB population dynamics simulations at $y / \beta = 1$}
\label{sec:Popdynyb1}

In this population dynamics, each individual $i$ is composed by four
probability functions: $\textbf{q0}_i$, $\textbf{qi}_i$, $\textbf{qX}_i$,
which is also the probability function in functional $Q^0_{i\rightarrow j}$,
$Q^i_{i\rightarrow j}$, $Q^X_{i\rightarrow j}$ correspondingly, 
and $\overline{\textbf{q}}_i$, which is the mean value of the probability $q_i$
over the distribution $Q[q]$. Here we present the pseudocode of the 1RSB
population dynamics at $y/ \beta =1$:
\begin{algorithm}[H]
  \algsetup{indent=1em}
  \caption{1RSB population dynamics at $m=1$}
  \begin{algorithmic}
    \FOR{$t=1, \cdots, T$}
    \FOR{$r=1, \cdots, N$}
    \STATE {Draw $k$ with cavity degree distribution}
    \STATE Draw $i(1), \cdots, i(k)$ uniformly in $\{1, \cdots, N\}$
    \newline
    \STATE Compute
    $\overline{\textbf{q}}=b(\overline{\textbf{q}}_{i(1)},\cdots,
    \overline{\textbf{q}}_{i(k)})$
    \newline
    \FOR{$s=1, \cdots, k$}
    \STATE {Draw $\textbf{q}_{i(s)}=\{\textbf{q0}_{i(s)},
      \textbf{qi}_{i(s)}, \textbf{qX}_{i(s)}\}$ with probability
      $\overline{q}_{i(s)}^0$, $\overline{q}_{i(s)}^{i(s)}$,
      $\overline{q}_{i(s)}^X$}
    \ENDFOR
    \STATE Compute $\textbf{q0}=b(\textbf{q}_{i(1)},\cdots,
    \textbf{q}_{i(k)})$
    \newline
    \FOR{$s=1, \cdots, k$}
    \STATE {Draw $\textbf{q}_{i(s)}=\{\textbf{q0}_{i(s)}, \textbf{qi}_{i(s)}\}$
      with probability $\frac{\overline{q}_{i(s)}^0}{\overline{q}_{i(s)}^0+
        \overline{q}_{i(s)}^{i(s)}}$, $\frac{\overline{q}_{i(s)}^{i(s)}}{
        \overline{q}_{i(s)}^0+\overline{q}_{i(s)}^{i(s)}}$}
    \ENDFOR
    \STATE Compute $\textbf{qi}=b(\textbf{q}_{i(1)},\cdots,
    \textbf{q}_{i(k)})$
    \newline
    \STATE {Draw $l$ with probability $w_l$}
    \FOR{$s=1, \cdots, k$}
    \IF {$s==l$}
    \STATE {Draw $\textbf{q}_{i(s)}=\{\textbf{qi}_{i(s)}, \textbf{qX}_{i(s)}\}$
      with probability $\frac{\overline{q}_{i(s)}^{i(s)}}{
        \overline{q}_{i(s)}^{i(s)}+\overline{q}_{i(s)}^X}$,
      $\frac{\overline{q}_{i(s)}^X}{\overline{q}_{i(s)}^{i(s)}+
        \overline{q}_{i(s)}^X}$}
    \ELSE
    \STATE {Draw $\textbf{q}_{i(s)}=\{\textbf{q0}_{i(s)}, \textbf{qi}_{i(s)}\}$
      with probability $\frac{\overline{q}_{i(s)}^0}{\overline{q}_{i(s)}^0+
        \overline{q}_{i(s)}^{i(s)}}$, $\frac{\overline{q}_{i(s)}^{i(s)}}{
        \overline{q}_{i(s)}^0+\overline{q}_{i(s)}^{i(s)}}$}
    \ENDIF
    \ENDFOR
    \STATE Compute $\textbf{qX}=b(\textbf{q}_{i(1)},
    \cdots,\textbf{q}_{i(k)})$
    \newline
    \STATE Draw $j$ uniformly in \{1, $\cdots$, N \}
    \STATE Save $\overline{\textbf{q}}$, $\textbf{q0}$, $\textbf{qi}$,
    $\textbf{qX}$ to individual $j$
    \ENDFOR
    \ENDFOR
    \label{code:recentEnd}
  \end{algorithmic}
\end{algorithm}

Here we use this population dynamics to compute the complexity $\Sigma$ of the
RR graph at the 1RSB region and the results are presented in
Fig.~\ref{fig:Complexity}. The size of the population for both graph ensembles
is $128k$. Before we record our results, we run 1RSB population dynamics $16k$
steps to relax the system. After that, we record and average the complexity of
the following $128k$ steps. To evaluate and  minimize the errors of our
computation, each point in Fig.~\ref{fig:Complexity} is the mean value of the
computation from 16 different random seeds. The error bar in 
Fig.~\ref{fig:Complexity} is the uncertainty of the mean value result.

\section{Population dynamics of the energetic 1RSB cavity method.}
\label{sec:appSPy}

In this algorithm, the whole population is composed by $N$ individuals, and 
each individual in the population is a distribution function of $\chi$, which 
is denoted by $\Theta_i(\chi)$ and is presented by $M$ cavity fields $\chi_j$.
Therefore, the algorithm will use a two-dimensional population structure and
we will use $\chi_{ij}$ to denote the $j$th cavity field in the $i$th
individual.
Usually, we are confined by the computation resource and cannot use a very
large $M$.
In that case, when $y$ is large, the cavity fields with large $\gamma$ will
dominate the result. However, the entropy of this kind of cavity field might
be very small. In order to make sure the algorithm can sample a typical cavity
field, we introduce a new algorithm parameter $N_p$ to enlarge the reweighting
space. Generally, we can use a small $N_p$ when $y$ is close to $0$ and then 
the value of $N_p$ should be raised with $y$ exponentially.
This parameter will increase the reweighting accuracy of the final result but
also increase the computation cost.
Here we present the pseudocode of this algorithm in a graph ensemble.
\begin{algorithm}[H]
  \algsetup{indent=1em}
  \caption{Population dynamics of the energetic 1RSB cavity method
    (size $N$, $M$, iterations $T$)}
  \begin{algorithmic}
    \STATE Initial the population $\{\Theta_i(\chi)\}$
    \FOR{$t=1, \cdots, T$}
    \FOR{$r=1, \cdots, N$}
    \STATE Draw $k$ with cavity degree distribution $\rho$
    \STATE Draw $i(1), \cdots, i(k)$ uniformly in $\{1, \cdots, N\}$
    \FOR{$s=1, \cdots, N_p\times M$}
    \STATE Draw $j(1), \cdots, j(k)$ uniformly in $\{1, \cdots, M\}$
    \STATE Compute $\chi_s=b_{SPy}\bigl(\{\chi_{i(k)j(k)}\}\bigr)$
    \STATE Compute $W_s=\exp(y\times\gamma_{i(k)j(k)})$
    \ENDFOR
    \STATE Generate the new population
    $\Theta_r(\chi)=\mathrm{REWEIGHT}(\{\chi_s,W_s\})$
    \ENDFOR
    \ENDFOR
  \end{algorithmic}
\end{algorithm}
The reweighting procedure is given by:
\begin{algorithm}[H]
  \algsetup{indent=1em}
  \caption{REWEIGHT(population of messages and weights $\{\chi_s,W_s\}$)}
  \begin{algorithmic}
    \FOR{$s=1, \cdots, N_p\times M$}
    \STATE Set $p_s\equiv W_s/\sum_s W_s$
    \ENDFOR
    \FOR{$s=1, \cdots, M$}
    \STATE Draw $i\in \{1, \cdots, N_p\times M\}$ with probability $p_s$
    \STATE Set $\chi^{new}_s=\chi_i$
    \ENDFOR
    \RETURN $\{\chi^{new}_s\}$.
  \end{algorithmic}
\end{algorithm}

\endwidetext


\end{document}